\documentclass[iop,twoside]{emulateapj}
\usepackage{graphicx,psfig,amssymb,float,natbib,rotating,txfonts}

\shorttitle{A {\em Chandra} X-ray study of NGC\,6791}
\shortauthors{van den Berg et al.}

\begin{document}

\title{A {\em Chandra} X-ray study of the interacting binaries in the old open cluster NGC\,6791} 

\author{Maureen van den Berg\altaffilmark{1,2}, Frank
  Verbunt\altaffilmark{3,4}, Gianpiero Tagliaferri\altaffilmark{5},
  Tomaso Belloni\altaffilmark{5}, Luigi R. Bedin\altaffilmark{6}, and
  Imants Platais\altaffilmark{7}}

\affil{\altaffilmark{1}Astronomical Institute ``Anton Pannekoek'',
  University of Amsterdam, Science Park 904, 1098 XH Amsterdam, The
  Netherlands}
\affil{\altaffilmark{2}Harvard-Smithsonian Center for Astrophysics,
    60 Garden Street, Cambridge, 02138 MA, USA; maureen@head.cfa.harvard.edu}
\affil{\altaffilmark{3}Department of Astrophysics/IMAPP, Radboud
  University Nijmegen, PO Box 9010, 6500 GL Nijmegen, The Netherlands}
\affil{\altaffilmark{4}SRON, Netherlands Institute for Space Research,
  Sorbonnelaan 2, 3584 CA Utrecht, The Netherlands}
\affil{\altaffilmark{5}INAF/Osservatorio Astronomico di Brera, Via
    E. Bianchi 46, I-23807 Merate (LC), Italy}
\affil{\altaffilmark{6}INAF/Osservatorio Astronomico di Padova,
Vicolo dell'Osservatorio 5, I-35122 Padova, Italy}
\affil{\altaffilmark{7}Department of Physics and Astronomy, The Johns
  Hopkins University, Baltimore, MD 21218, USA}

\hyphenation{sub-dwarf}

\begin{abstract}
We present the first X-ray study of NGC\,6791, one of the oldest open
clusters known (8 Gyr). Our {\em Chandra} observation is aimed at
uncovering the population of close interacting binaries down to
$L_{\rm X} \approx 1\times10^{30}$ erg s$^{-1}$ (0.3--7 keV). We
detect 86 sources within 8\arcmin~of the cluster center, including 59
inside the half-mass radius. We identify twenty sources with
proper-motion cluster members, which are a mix of cataclysmic
variables (CVs), active binaries (ABs), and binaries containing
sub-subgiants. With follow-up optical spectroscopy we confirm the
nature of one CV. We discover one new, X-ray variable candidate CV
with Balmer and He~II emission lines in its optical spectrum; this is
the first X-ray--selected CV confirmed in an open cluster. The number
of CVs per unit mass is consistent with the field, suggesting that the
3--4 CVs observed in NGC\,6791 are primordial. We compare the X-ray
properties of NGC\,6791 with those of a few old open (NGC\,6819,
M\,67) and globular clusters (47\,Tuc, NGC\,6397).  It is puzzling
that the number of ABs brighter than $1\times10^{30}$ erg s$^{-1}$
normalized by cluster mass is lower in NGC\,6791 than in M\,67 by a
factor $\sim$3--7. CVs, ABs, and sub-subgiants brighter than
$1\times10^{30}$ erg s$^{-1}$ are under-represented per unit mass in
the globular clusters compared to the oldest open clusters, and this
accounts for the lower total X-ray luminosity per unit mass of the
former. This indicates that the net effect of dynamical encounters may
be the destruction of even some of the hardest (i.e.~X-ray--emitting)
binaries.
\end{abstract}


\section{Introduction}

X-ray emission of late-type stars arises in hot gas contained by
magnetic loops above the stellar surface. These loops are produced by
the interaction of convective motion and differential rotation in the
stellar envelope; hence the magnetic activity of a late-type
main-sequence star---and with it the X-ray emission---is found to be
higher in rapidly rotating stars \citep[see, for example,][]{pall89}.
As stars age their rotation slows down, and it was a surprise that a
{\em ROSAT} study of the old open cluster M\,67 discovered a large
number of X-ray sources \citep{bellea93}.  Comparison with optical
observations provided the explanation: in an old cluster stars may
rotate rapidly when tidal forces spin them up towards corotation with
the binary revolution \citep{bellea}.  X-ray studies of old open
clusters help in identifying such, otherwise inconspicuous, active
binaries (ABs) in which tidal forces are or have been effective, and
contribute to the understanding of tidal interaction. Since the
typical timescale for tidal interaction rapidly increases with the
ratio of the orbital separation to stellar radius \citep{zahn89}, only
stars in binaries with relatively short periods or relatively large
radii experience the effects of tidal locking. X-ray observations of
ABs in old clusters thus probe the populations of hard binaries, and
the relative number of such systems found in clusters with different
properties is an important tool for understanding the effects of the
cluster environment on the binary content.

Thanks to prolonged optical studies, e.g.~in the WIYN Open Cluster
Study \citep{math00}, our knowledge of the binary population in
clusters is reaching completeness levels which enable a detailed
comparison with theoretical models of the evolution of stellar
clusters and of the binary population in them. To explain the large
number of blue stragglers observed in M\,67 (4 Gyr), the model by
\cite{hurlea05} requires a large initial number of short-period
binaries that act as seeds from which blue stragglers are formed. As
the simulated hard-binary fraction and period distribution do not
evolve much, the close binaries remain abundant throughout the life
time of the cluster. \cite{gellea12} point out that comparison with
the observed binaries in the old cluster NGC\,188 (6.5 Gyr;
\citealt{meibea09}) finds no evidence of such a high fraction of close
binaries. The number of clusters studied is still small, as is the
number of detailed models, and more study is required before a final
verdict on these and other discrepancies (see \cite{gellea12} for
details) can be made.

By now, observations of globular clusters with the {\em Chandra X-ray
  Observatory} have uncovered hundreds of faint ($L_{\rm X} \lesssim
10^{33}$ erg s$^{-1}$) close binaries, only part of which have been
classified so far. As these binaries are an important driver of the
long-term evolution of a cluster, understanding the properties and
frequency of the various binary classes is a major goal of
investigating (globular) cluster X-ray sources. Sometimes the outcome
of a dynamical encounter is a binary or multiple system in a very
unusual configuration, and this allows us to study source types that
are not, or only rarely, found in the field. Besides ABs, which are
the dominant X-ray source class in old open clusters, globular-cluster
X-ray sources include cataclysmic variables (CVs), low-mass X-ray
binaries in quiescence (qLMXBs), and milli-second pulsars (MSPs). The
correlation of the numbers of qLMXBs and CVs with the stellar
encounter frequency reveals the important role that dynamical
encounters play in the creation of such binaries
\citep{poolhut06,heinea06}, but the net balance between formation and
destruction is less clear. For a large part this is due to the lack of
a reference point that quantifies their number densities in a
dynamically inactive environment. Ideally, the comparison should be
made against the Galactic field where stellar densities are low, but
this is complicated because of the generally limited information on
distance and age for stars in the field, and the intrinsic scarcity of
sources such as qLMXBs. Old open clusters provide an alternative that
is worthwhile to explore, but so far only few have been studied in
detail in X-rays. Moreover, it should be kept in mind that dynamical
encounters cannot be completely ignored in open
clusters. Observational indications for the occurrence of dynamical
interactions have come from individual systems whose properties are
difficult to explain by binary evolution as it would take place
outside a cluster environment \citep[e.g.][]{vdbergea2001ad}.

In this paper we present the results of the first X-ray observation of
NGC\,6791, which, at an age of about 8 Gyr, is one of the oldest known
open clusters in our Galaxy. Besides the {\em Chandra} data, we
obtained optical spectra to classify candidate optical counterparts to
the detected X-ray sources. NGC\,6791 has been studied extensively in
the optical. The rich body of available literature has proven to be
very useful for the classification of our {\em Chandra} sources, and
for separating cluster members from non-members.  For the study of
interacting binaries it is a promising target: before this work it was
known to harbor two of only three spectroscopically-confirmed CVs
found in open clusters \citep{kaluea97}, and many optical-variability
studies have revealed dozens of close binaries in the field of the
NGC\,6791 (see \cite{demaea07} and references therein). It is
therefore interesting to investigate if the X-ray source content of
NGC\,6791 is as rich and varied as was found for the few old open
clusters studied previously, viz.~M\,67
\citep{bellea93,bellea,vdbergea04} and NGC\,188
\citep{bellea,gond05}. The cluster lies at a distance of
$\sim$4.0--4.3 kpc and is reddened by $E(B-V)$=0.09--0.16; see
e.g.~\cite{carrea06}, \cite{basuea11}, and \cite{brogea11} for recent
determinations of the cluster parameters. Throughout the paper we
adopt a distance of 4.1 kpc and a constant reddening $E(B-V)=0.14$,
which corresponds to a neutral-hydrogen column density $N_{\rm H}=7.8
\times 10^{20}$ cm$^{-2}$ \citep{predschm95}. The reddening could vary
across the area of the cluster \citep{platea11,brogea12} but the
effect is too small to have a significant impact on our
results. NGC\,6791 stands out in several ways, including its high age
and large mass. Remarkably for such an old cluster, the metallicity is
higher than solar ([Fe/H]$\approx$+0.4).

The X-ray and optical observations and data reduction are described in
Sect.~\ref{sec_obs}, followed by the data analysis in
Sect.~\ref{sec_ana}. We detected various types of sources in the
cluster which are presented in Sect.~\ref{sec_results}. Our results
are discussed in the context of the populations of interacting
binaries in star clusters in Sect.~\ref{sec_disc}. We finish with our
conclusions in Sect.~\ref{sec_conc}.

\section{Observations and data reduction} \label{sec_obs}

\subsection{X-ray observations}

\begin{figure}
\centerline{\includegraphics[width=8.3cm]{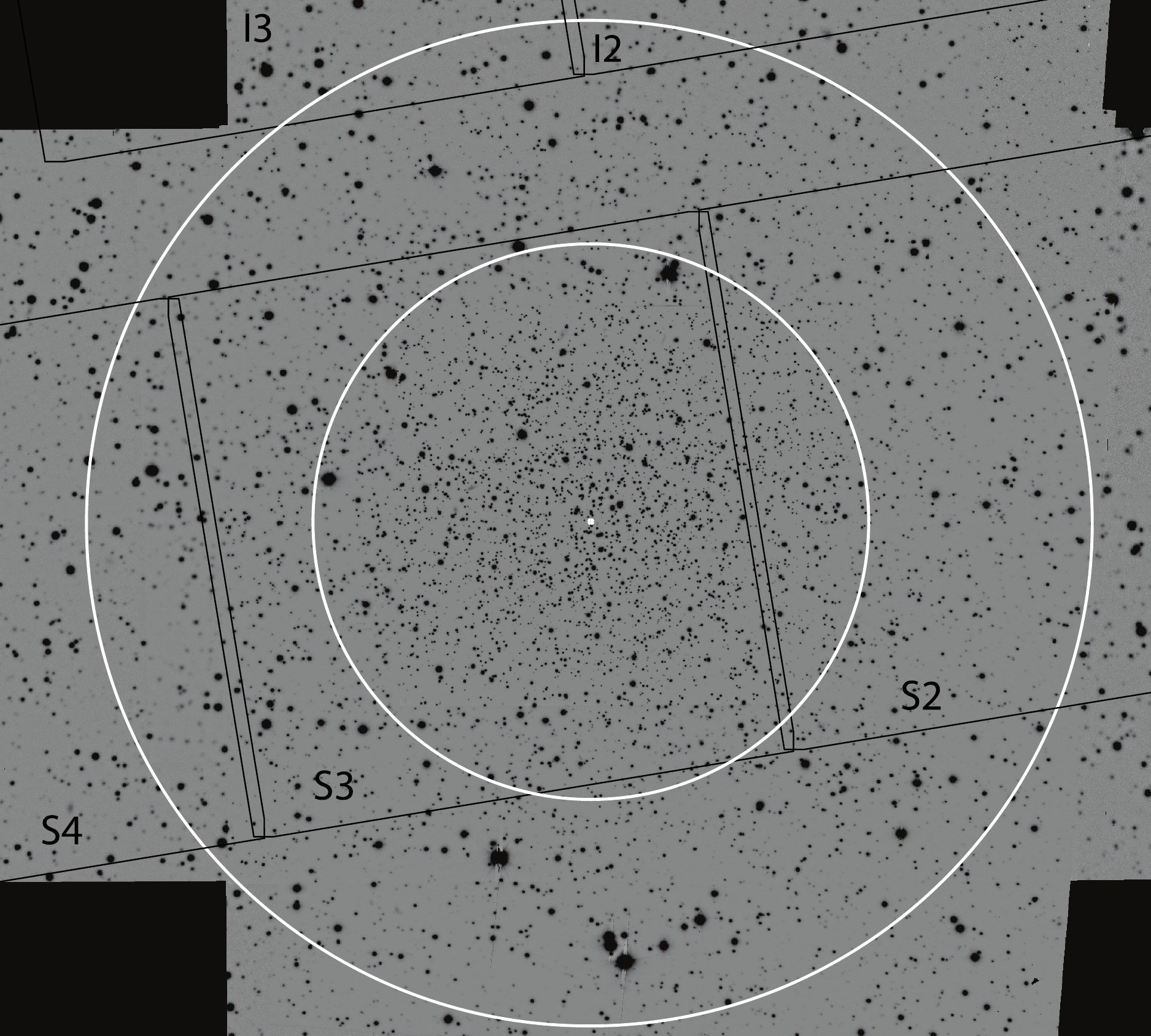}}
\caption{$V$-band image of NGC\,6791 from \cite{stetea03} showing in
  black the outline of the ACIS chips (labelled) as in our {\em
    Chandra} observation. The small white, innermost circle indicates
  the cluster center from \cite{demaea07}. The larger white circles
  mark the area inside the half-mass radius $r_h$ (4\farcm42), and the
  area inside 8\arcmin, which was analyzed in this paper. North is up,
  east to the left. \label{fig_fov}}
\end{figure}

We observed NGC\,6791 with the Advanced CCD Imaging Spectrometer
(ACIS)\footnote{http://cxc.harvard.edu/proposer/POG/html/ACIS.html} on
{\em Chandra} from 2004 July 1 20:51 UTC until July 2 10:49 UTC for a
total exposure time of 48.2 ks (ObsID 4510). The observation was done
in very faint, timed exposure mode with a single-frame exposure time
of 3.2 s. In order to achieve optimal sensitivity below $\sim$1 keV,
we placed the central part of the cluster on the backside-illuminated
S\,3 chip; the neighboring S\,1, S\,2, S\,4, I\,2, and I\,3 chips were
used to cover the outer parts of the cluster. The cluster center lies
at approximately $\alpha$=19$^{\rm h}$20$^{\rm m}$53$^{\rm s}$,
$\delta$=+37$^{\circ}$46\arcmin18\arcsec~(J2000; \citealt{demaea07});
we shifted the center 0\farcm5 away from the S\,3 aimpoint in the $-Y$
(approximately south-east) direction so that a larger part of the
cluster could be imaged on a single chip. \cite{platea11} derive a
cluster half-mass radius $r_h$ of $4\farcm42\pm0\farcm02$. Our
observation covers almost the entire area inside $r_h$
($\gtrsim98.5$\%) with either the S\,3 or the S\,2 chip; only the
southernmost tip falls off the detector area. Fig.~\ref{fig_fov} shows
the outline of the ACIS chips overlayed on an optical image of the
cluster.

Our data reduction started with the level-1 event file produced by
version 7.6.7.2 of the {\em Chandra} X-ray Center processing pipeline.
We used the CIAO 3.4 package with the CALDB 3.3.0 calibration files
for further processing following standard
procedures\footnote{http://cxc.cfa.harvard.edu/ciao3.4/guides/acis\_data.html},
including the very-faint mode background cleaning.  In order to
improve the precision of the source positions we removed the
randomization of event coordinates that is applied in standard
processing.  Towards the end of the observation, the background level
between 0.3 and 7.0 keV increased by a factor of $\sim$5 as a result
of a background flare. After removing this interval from the
observation, the exposure time is reduced to 42.9 ks.

\subsection{Optical spectroscopy} \label{sec_specid}

We obtained low-resolution spectra of candidate optical counterparts
to guide the classification of the X-ray sources. A total of 16
candidate counterparts brighter than $V\approx18.3$ were observed with
the FAST long-slit spectrograph on the 1.5-m Tillinghast telescope on
Mt. Hopkins on 9 nights between 2005 June 7 to September 2. We used
the 300 lines mm$^{-1}$ grating, resulting in a wavelength coverage
from 3480 to 7400 \AA\,\,and a 3-\AA~resolution. Exposure time was
chosen to achieve a signal-to-noise ratio $S/N$ $\ga20$ for
$V\la17.5$, and $10 \la S/N \la 20$ for fainter sources. FAST spectra
were extracted and wavelength-calibrated with a dedicated reduction
pipeline\footnote{http://tdc-www.harvard.edu/instruments/fast}.  Flux
standards were observed on the same nights as the science targets.

Candidate optical counterparts fainter than $V\approx17$ were observed
with the fiber-fed multi-object spectrograph Hectospec on the 6.5-m
Multi-Mirror Telescope. Use of the 270 lines mm$^{-1}$ grating
resulted in spectra that cover 3700 to 9150 \AA\,\,with a resolution
of 6 \AA. A total of 16 candidate counterparts were observed on the
nights of 2005 May 13 and July 4--6. Each setup was repeated 4 to 5
times with individual exposures of 900 s. Exposures were combined, and
the spectra extracted and wavelength-calibrated, with the Hectospec
reduction pipeline\footnote{
  http://tdc-www.harvard.edu/instruments/hectospec/}. Sky background
spectra for each setup were created by combining the spectra of fibers
positioned at off-source locations.  Hectospec observations did not
include flux standards. To correct for the instrumental response we
constructed sensitivity curves using the spectra of the subdwarf B
(sdB) stars B\,4, B\,5 and B\,3 in NGC\,6791 \citep{kaluudal92},
included in the May 13, July 4 and July 5--6 setups, respectively.  We
assumed that their intrinsic spectra can be described by blackbodies
with temperatures as determined for these sdB stars by
\cite{liebea94}. The blackbody spectra were then normalized to
reproduce the observed $V$ magnitudes (\citealt{stetea03}) after
correction for the cluster reddening. This way we achieved an
approximate absolute-flux calibration of the target spectra after
applying the resulting sensitivity curves.

Flux-calibrated target spectra were assigned spectral types by
comparison to standard spectra with similar resolution
\citep[e.g.][]{jacoea84}.

\section{Analysis} \label{sec_ana}

\subsection{X-ray source detection and extraction} \label{sec_xdet}

\begin{table*}
\caption{{\em Chandra} sources within 8\arcmin~of the center of NGC\,6791 \label{tab_src}}
\begin{center}
\begin{tabular}{lccccccccc}
\hline
\hline
\multicolumn{1}{c}{(1)} & (2)     & (3)      & (4)      & (5)     & (6)     & (7)    & (8)                  & (9)      & (10) \\ 
CX     & CXOU\,J & $\alpha$(J2000) & $\delta$(J2000) & $r_{95}$ & Offset  & Counts & $F_{\rm X,u}$          & $E_{50}$ & Opt \\
                        &         & (deg)  & (deg)  & (\arcsec) & (\arcmin) &        & (erg cm$^{-2}$ s$^{-1}$) & (keV)     &     \\
\hline\\[-8pt]
\multicolumn{10}{c}{Sources inside the half-mass radius $r_h$}\\[3pt]
\hline
~~1  & 192044.9+374640 & 290.187175 & +37.777870 & 0.32 & 1.64 & 213 $\pm$ 15 & 318.0 & 1.3 $\pm$ 0.1   &     + \\ 
~~2  & 192039.8+374354 & 290.165914 & +37.731752 & 0.39 & 3.54 & 151 $\pm$ 13 & 242.9 & 1.00 $\pm$ 0.03 &     + \\ 
~~3  & 192035.7+374452 & 290.148996 & +37.747852 & 0.39 & 3.70 & 125 $\pm$ 12 & 260.1 & 1.4 $\pm$ 0.1   &     + \\ 
~~4  & 192056.3+374613 & 290.234628 & +37.770331 & 0.34 & 0.66 & 116 $\pm$ 11 & 170.7 & 1.4 $\pm$ 0.1   &     + \\ 
~~5  & 192047.3+374318 & 290.197161 & +37.721768 & 0.43 & 3.20 &  72 $\pm$  9 & 118.0 & 1.4 $\pm$ 0.1   &   $-$ \\ 
~~6  & 192114.4+374530 & 290.310342 & +37.758603 & 0.63 & 4.32 &  72 $\pm$  9 & 112.0 & 1.4 $\pm$ 0.1   &   $-$ \\ 
~~7  & 192052.3+374550 & 290.218110 & +37.764148 & 0.35 & 0.47 &  53 $\pm$  8 &  77.9 & 1.2 $\pm$ 0.1   &     + \\ 
~~8  & 192038.2+374441 & 290.159518 & +37.744838 & 0.46 & 3.33 &  48 $\pm$  7\tablenotemark{b} & 221.7 & 1.6 $\pm$ 0.2   &     + \\ 
~~9  & 192058.4+375008 & 290.243732 & +37.835590 & 0.56 & 3.98 &  37 $\pm$  7 &  59.8 & 1.6 $\pm$ 0.3   &     + \\ 
10  & 192037.3+374612 & 290.155726 & +37.770073 & 0.43 & 3.09 &  34 $\pm$  6 &  69.3 & 1.4 $\pm$ 0.2   &   $-$ \\ 
\hline
\end{tabular}

\tablecomments{Columns: 1) source number; 2) source name; 3 and 4)
  source coordinates in decimal degrees after applying the boresight
  correction of $\Delta\alpha=-0\farcs06\pm0\farcs06$, $\Delta\delta=
  -0\farcs21\pm0\farcs04$ ({\em Chandra} minus optical); 5) 95\%
  uncertainty radius on the source position; 6) angular offset from
  the cluster center; 7) net counts in the 0.3--7 keV band; 8)
  unabsorbed flux ($\times~10^{-16}$ erg s$^{-1}$ cm$^{-2}$) in the
  0.3--7 keV band for the assumption that the source spectrum is a
  2-keV Mekal plasma seen through a neutral-hydrogen column of density
  $7.8 \times 10^{20}$ cm$^{-2}$; 9) median energy in the 0.3--7 keV
  band (only for sources with $>$5 counts); 10) flag for the detection
  of a candidate optical counterpart. Table~\ref{tab_src} is available
  in its entirety in a machine-readable form in the online journal. A
  portion is shown here for guidance regarding its form and content.}

\end{center}
\end{table*}

We restrict the analysis to the area within 8\arcmin\, of the cluster
center. Sources further away have relatively large positional errors,
which complicates the optical identification. We focus on the area
inside $r_h$ which is close to the largest circular area around the
cluster center that is entirely covered by the observation.

We performed source detection in a broad (0.3--7.0 keV), soft
(0.3--2.0 keV) and hard (2.0--7.0 keV) energy band, also used in our
{\em Chandra} study of M\,67 \citep{vdbergea04}, to facilitate
comparison. The CIAO detection routine {\em wavdetect} was run for
scales of 1.0 to 11.3 pixels, in steps increasing by a factor
$\sqrt{2}$, with the larger scales appropriate for large off-axis
angles where the point-spread function (PSF) becomes significantly
broader.  We computed exposure maps for the response at 1 keV to
account for spatial variations of the sensitivity. The {\em wavdetect}
detection threshold was set to 10$^{-6}$, from which we expect two
spurious detections per detection scale (so sixteen spurious
detections in total) in the area that we consider here. {\em
  Wavdetect} positional errors reflect the statistical uncertainty in
centroiding the spatial distribution of the detected events of a given
source, but do not include systematic errors that result from data
processing and PSF asymmetries at large offset angles. To compute
positional uncertainties we therefore adopt Eq.~5 in
\cite{hongvandea05}. This formula relates the 95\% confidence radius
on the source position $r_{95}$ to the {\em wavdetect} counts and
offset angle from the aimpoint, and is based on extensive simulated
detections of artificial sources.  Combination of the broad, soft, and
hard-band source lists results in a master catalog of 86 distinct
sources within 8\arcmin~of the cluster center, of which 59 lie inside
$r_h$. Table~\ref{tab_src} summarizes their basic properties. To
investigate the validity of the sources, we also ran {\em wavdetect}
with a threshold of 10$^{-7}$ or an expected number of spurious
sources of 1.6. The fourteen sources not detected in this run are
marked with an asterisk in Table~\ref{tab_src}.  In this paper we
adopt a short version of the source names to refer to the sources
instead of using their official CXO names; see columns 1 and 2 in
Table~\ref{tab_src}. The shorthand names are assigned by first sorting
the sources within $r_h$ on net counts, and then the sources between
$r_h$ and 8\arcmin.

We used the {\em acis\_extract} package \citep[version 3.107.2;
][]{brooea02} to measure net source counts. Events between 0.3 and 7.0
keV were extracted from a region corresponding to 90\% of the PSF at
1.5 keV; for a few sources this region was reduced to avoid
contamination by a close neighbor. The background was determined from
a source-free annulus centered on the source position. We convert net
count rates to fluxes within {\tt
  Sherpa}\footnote{http://cxc.harvard.edu/sherpa} with arf and rmf
response files appropriate for the chip location and source-extraction
area of each source.  Column 7 of Table~\ref{tab_src} lists
absorption-corrected ($u$) fluxes $F_{\rm X,u}$ computed under the
assumption that the underlying spectrum is an optically thin plasma
(described by the {\em xsmekal} model) with $kT=2$ keV.  For source
CX\,29 that lies near the aimpoint 1 count s$^{-1}$ corresponds to
$6.8 \times 10^{-12}$ erg cm$^{-2}$ s$^{-1}$. This temperature is
appropriate for the most active ABs in the cluster (see
e.g.~\citealt{vdbergea04}), while too high for the least active ABs
and too low for most CVs. For a 1-keV Mekal model, a power-law
spectrum ({\em powlaw1d}) with photon index $\Gamma=2$, and a 10-keV
thermal-bremsstrahlung model ({\em xsbremss}) the conversion factor is
24\% smaller, and 30\% and 42\% larger, respectively. In each case we
account for a column density equal to the cluster value using the {\em
  xsphabs} model. For the adopted distance to NGC\,6791 of 4.1 kpc, a
3-count detection limit corresponds to a limiting unabsorbed X-ray
luminosity $L_{\rm X,u}$ of $(0.7-1.4)\times 10^{30}$ erg s$^{-1}$
cm$^{-2}$ (0.3--7 keV) where the range corresponds to the choice of
models specified above.

{\em Acis\_extract} performs a Kolmogorov-Smirnov test on the event
arrival times to test for variability. Two sources are thus found to
be potentially variable, with probabilities that their count rates are
constant ($p_{\rm K-S}$) smaller than 2.5\%: CX\,19 (a candidate CV,
see Sect.~\ref{sec_cv}) for which 15 of a total of 19 counts are
detected within 5.5 hours ($p_{\rm K-S}=0.018$), and CX\,37 (no
optical counterpart found) for which all 9 counts are detected in the
first 7 hours of the observation ($p_{\rm K-S}=0.024$).

\subsection{X-ray spectral properties} \label{sec_xspec}

Only CX\,1 has sufficient counts to allow a constraining spectral fit.
A spectrum was extracted from the {\em acis\_extract} source region
with the CIAO tool {\em psextract}, and was grouped to have at least
20 counts bin$^{-1}$ to warrant use of the $\chi^2$ statistic; the
background contribution ($<$1 count) can be ignored. Given the
classification of CX\,1 as an active galactic nucleus (AGN) based on
its optical spectrum (Sect.~\ref{sec_galaxies}), we try to fit the
data with an absorbed power-law and find an acceptable result for a
photon index $\Gamma=1.9\pm0.3$ and a column density $N_{\rm H} < 1.2
\times 10^{21}$ cm$^{-2}$ (1-$\sigma$ upper limit; $\chi^2=9.5$, 6
degrees of freedom).  The limit on $N_{\rm H}$ is consistent with the
integrated Galactic column density in the direction of NGC\,6791
\citep[$9 \times 10^{20}$ cm$^{-2}$,][]{schlea98} while the value for
$\Gamma$ is typical for an AGN \citep[e.g.][]{tozzea06}.

The remaining sources have too few counts for useful constraints by
spectral fitting. Instead we use the method of quantile analysis
developed by \cite{hongschlea04}, where the median energy ($E_{50}$)
and the 25\% and 75\% energy quartiles ($E_{25}$ and $E_{75}$) of the
source photons are used to characterize spectral properties. The
advantage of using energy quantiles as opposed to comparing counts in
pre-defined energy bands by means of hardness ratios is that the
errors on the diagnostics are less sensitive to the underlying
spectral shape

\subsection{Optical identification} \label{sec_oid}

\subsubsection{Cross-identification against the Stetson catalog}

\nocite{stetea03} We looked for optical counterparts in the deep $BVI$
photometric catalog of NGC\,6791 compiled by Stetson et al.~(2003; S03
hereafter), which covers the entire area studied here. The limiting
magnitude is $V\approx24$ for the central area but the sensitivity is
lower for the outer regions. In an attempt to eliminate artifacts from
the catalog, we removed entries with photometric-quality indicators
that flag them as suspicious, viz.~sources with $|$sharpness$|$ $> 1$,
and separation index $< 0$; see S03 for an explanation of these
indicators and a motivation for these criteria.

We first measure and correct for the boresight, i.e.~a possible
systematic offset between the astrometric reference frames of the {\em
  Chandra} and optical positions. The optical positions are tied to
the International Celestial Reference System (ICRS) with an rms
accuracy of about 0\farcs27 (S03), but the astrometric reference frame
of any given {\em Chandra} observation as a whole is aligned to the
ICRS with a typical 0\farcs6 accuracy (90\%
uncertainty\footnote{http://cxc.harvard.edu/cal/ASPECT/celmon/}). We
measure the boresight correction using matches with optical variables
only. This minimizes the number of chance alignments as the light
curves of many variables reveal them as close binaries and therefore
as plausible X-ray emitters. A list of variables was compiled from the
studies of \cite{kaluruci93}, \cite{ruciea96},
\cite{mochea02,mochea03,mochea05}, \cite{brunea03}, \cite{hartea05},
and \cite{demaea07}; we only selected variables that we could identify
with objects in the S03 catalog so that we could use the S03 positions
for the boresighting. The boresight correction is computed
iteratively. We looked for matches to X-ray sources with more than 10
net counts and $r_{95} \leq 1\arcsec$. For each source, the match
radius is set to be the quadratic sum of the error on its X-ray
position, and the typical 1-$\sigma$ error on the optical positions
(0\farcs27) scaled to a 95\% error radius assuming a 2D gaussian error
distribution. The boresight is set to be the weighted (with
$r_{95}^{-2}$) mean of the X-ray--optical offsets of all matches
found. Next, the X-ray positions are corrected for the boresight, and
the matching is repeated, now including the statistical error on the
mean boresight in the match radius. This procedure quickly converges
to a boresight correction of $\Delta\alpha=-0\farcs06\pm0\farcs06$,
$\Delta\delta= -0\farcs21\pm0\farcs04$ ({\em Chandra} minus optical)
based on ten matches with optical variables. Matching the
boresight-corrected X-ray catalog to the clean S03 catalog results in
51 optical matches (including 26 variables) for 47 sources, out of the
86 X-ray sources in total.

The probability to find a match in the S03 catalog by chance depends
on the projected star density (which decreases with distance from the
cluster center $r$) and on the match radius (which typically increases
with $r$ because $r_{95}$ increases). To estimate the expected number
of spurious matches, we divide the cluster in a central area defined
by $r \leq r_h$, and an outer area of $r_h < r \leq 8\arcmin$. The
clean S03 catalog gives a projected density of 0.030 optical sources
arcsec$^{-2}$ and 0.0092 sources arcsec$^{-2}$ for the central and
outer area, respectively. If we multiply this by the total area
covered by the match circles of the X-ray sources, we find that the
expected number of random matches is 5.9 (center), and 3.8 (outer
annulus). This amounts to 15\% and 29\% of the matches found in the
respective regions. On the other hand, a similar calculation shows
that all matches with variable stars are likely to be real, with the
average number of chance alignments $<0.1$ both for the central and
outer regions.

In order not to overlook any matches, we repeated the matching using
the entire S03 catalog, and inspected the regions around the {\em
  Chandra} sources in the optical fits image from S03 to discard
matches with image artifacts or dubious detections. We thus found nine
extra candidate counterparts for eight {\em Chandra} sources, which
are not included in the clean catalog because the values of the
quality flags slightly exceeded the adopted cutoff limits, the object
is really extended, or because the object is faint and lies close to a
relatively bright star. Since selection of these additional
counterparts was not done in a very systematic way but relies on
visual inspection, it is not trivial to estimate the number of random
coincidences among the new matches. Two new candidate counterparts are
matched to a single {\em Chandra} source outside $r_h$, so at least
one of the two must be spurious. The other seven are uniquely matched
to seven {\em Chandra} sources inside $r_h$. Based on the higher
source density in the entire S03 catalog compared to the clean S03
catalog, one would expect to find $\sim$1.5 extra spurious matches in
the central area. Keeping in mind that some of these optical
``sources'' are not real, we estimate that at most one or two of the
seven new matches inside $r_h$ are spurious.

In total the S03 optical survey provides 60 astrometric matches for 55
sources.

\subsubsection{HST imaging} \label{sec_hst}

\begin{figure}
\centering
\includegraphics[width=8.3cm]{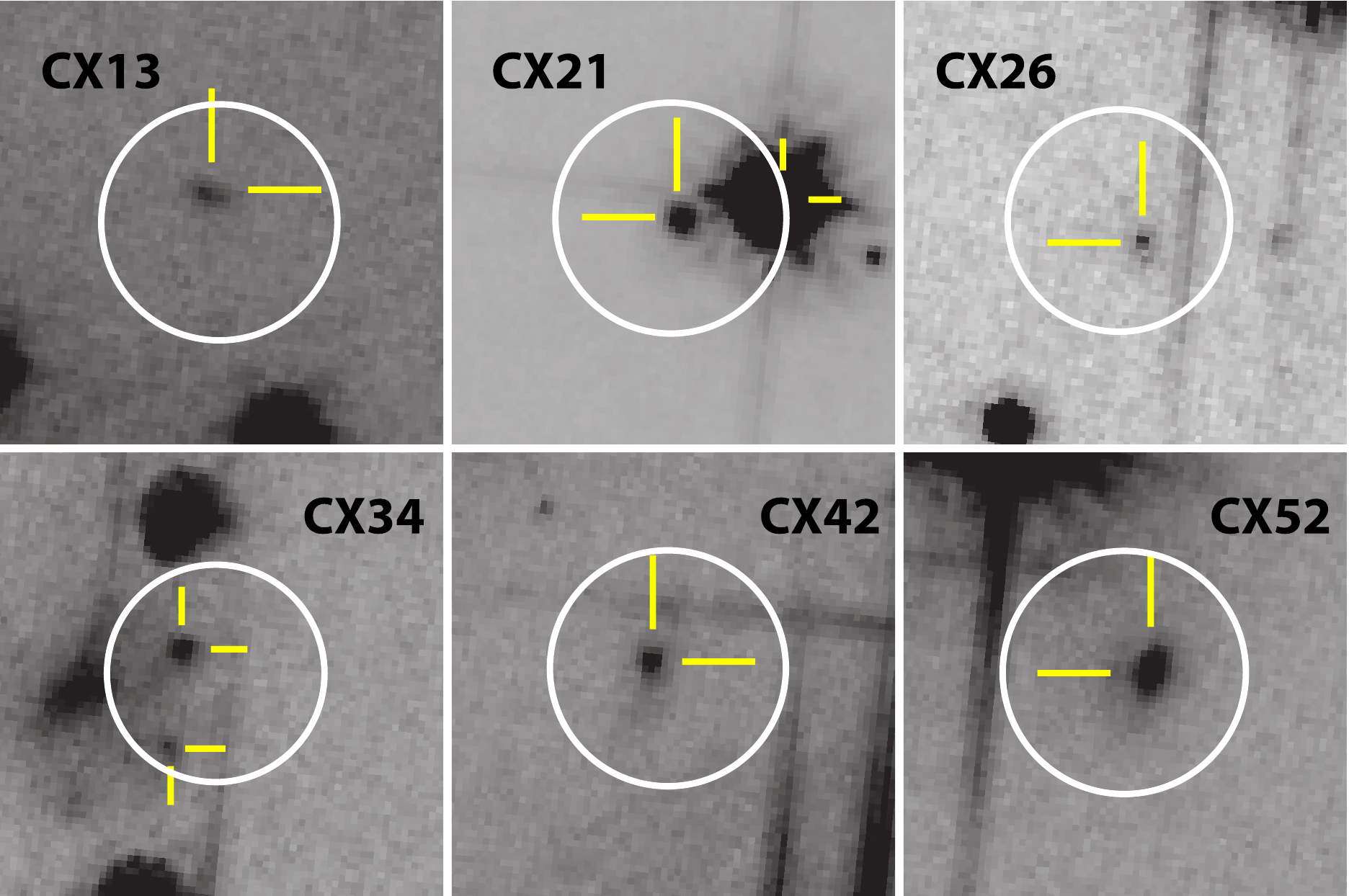}
\caption{ACS/WFC finding charts, each measuring 3\farcs4 $\times$
  3\farcs4~in size, made from the stacked GO-9815 F814W image
  representing 7024 s of exposure time in total.  The combined
  X-ray/optical 95\% error circles are shown as circles centered
  around the boresighted X-ray positions, while tick marks indicate
  the optical matches. Except for the bright candidate counterpart to
  CX\,21 on the edge of the error circle, these matches are found in
  the {\em HST} data only, and not in the \cite{stetea03} catalog.
  North is up, east to the left. \label{fig_hstid}}
\end{figure}

\begin{figure*}
\centerline{\includegraphics[width=17.5cm,clip]{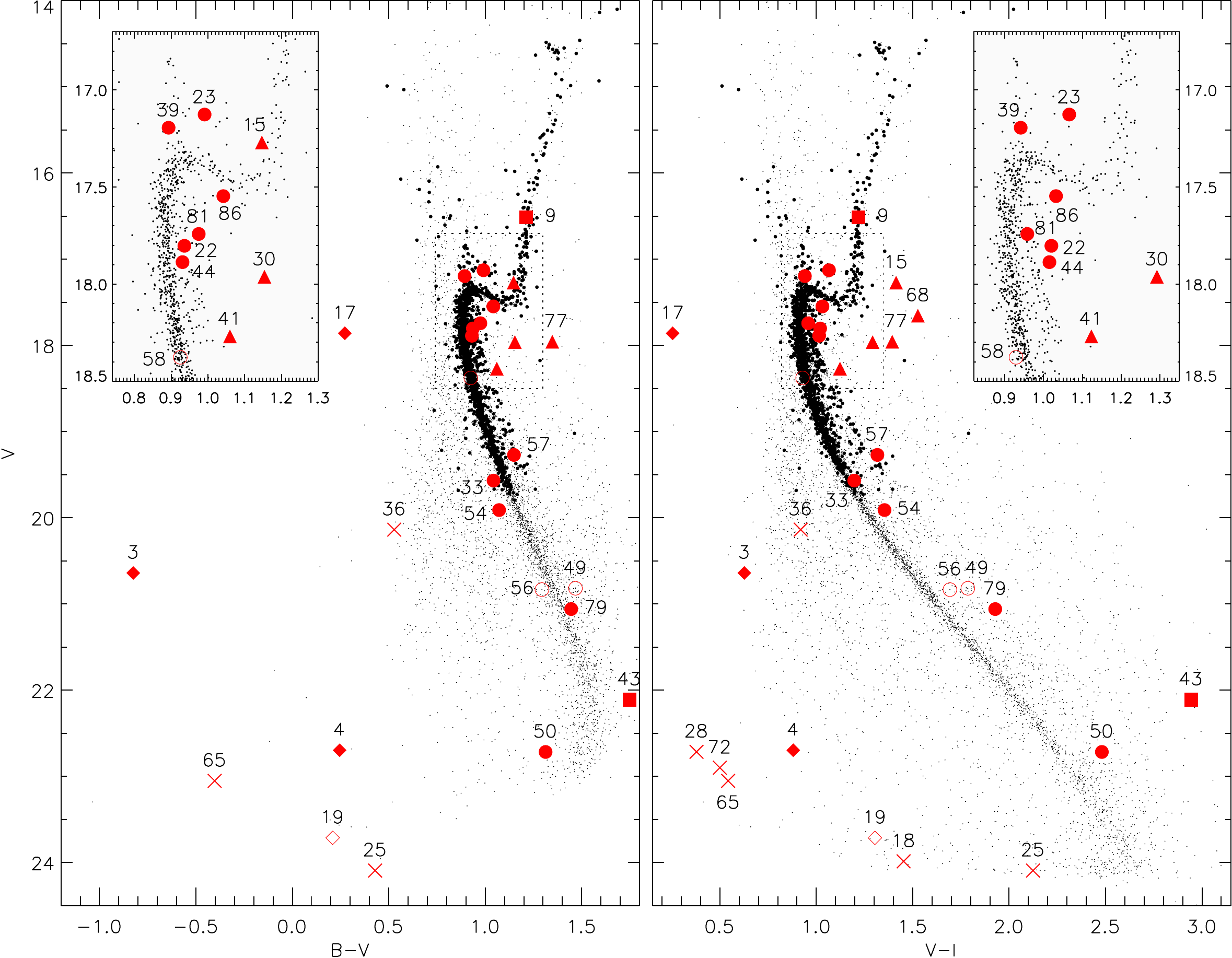}}
\caption{($V$, $B-V$) and ($V$, $V-I$) color-magnitude diagrams of
  NGC\,6791 with the candidate optical counterparts that are
  associated, or possibly associated, with the cluster marked in
  red. Candidate counterparts that are proper-motion members are
  indicated with large filled symbols, and sources without membership
  information with large open symbols or crosses: diamonds are
  (candidate) CVs, circles are (candidate) ABs, triangles are
  sub-subgiants, squares are stars without any indication of binarity,
  and crosses mark unclassified faint sources away from the main
  sequence. Stars in the field of NGC\,6791 are plotted as dots, while
  stars with a proper-motion membership probability $>50$\%
  (K. Cudworth, private communication) are shown as small filled black
  circles. The insets zoom in on the crowded regions of the
  diagrams. Photometry is from \cite{stetea03}. See the electronic
  edition of the Journal for a color version of this
  figure.  \label{fig_cmd_mem}}
\end{figure*}

\begin{figure}
\centering
\includegraphics[width=8.cm]{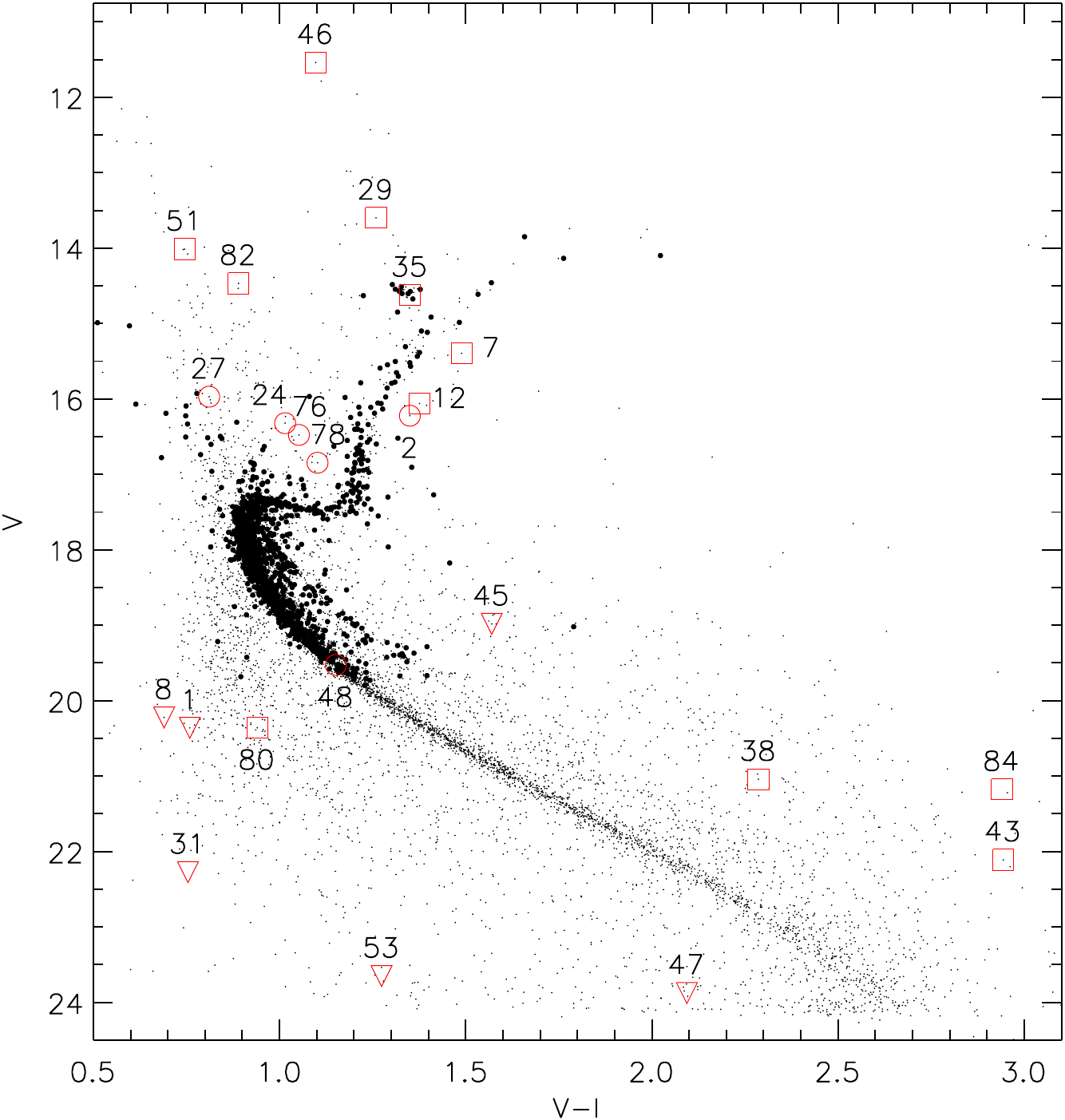}
\caption{Color-magnitude diagram of NGC\,6791 with candidate
  counterparts that are likely not associated with the cluster marked
  in red. Downward-pointing triangles are background galaxies, circles
  are (candidate) ABs, and squares are likely foreground stars without
  any indication of binarity. {\em HST} counterparts, for which
  photometry in $V$ and $I$ is not available, are not shown. See the
  electronic edition of the Journal for a color version of this
  figure. \label{fig_cmd_nmem}}.
\end{figure}

\cite{bediea06} used {\em Hubble Space Telescope} ({\em HST})
multi-epoch imaging with the Wide Field Channel on the Advanced Camera
for Surveys (ACS) to measure proper motions of objects in a central
3\farcm4 $\times$ 3\farcm4 region of NGC\,6791. We make use of their
results to establish cluster membership of candidate counterparts
included in the ACS images (Sect.~\ref{sec_membership}), and refer to
\cite{bediea06} for details of the data and the proper-motion
analysis. We also use these deep images to look for additional faint
candidate counterparts. The data set consists of F606W and F814W
images taken on 2003 July 16 and 17 (GO-9815), and on 2005 Jul 13
(GO-10471). An astrometric reference frame was created by combining
the GO-9815 F814W images with the STScI {\em Multidrizzle} software,
which removes artifacts like cosmic rays and bad pixels, and corrects
for the geometric distortion of ACS images.  In order to align the
coordinate system to the ICRS, we computed an astrometric solution
based on the positions of 273 unsaturated stars from the S03 catalog
that could be identified in the stacked image.  Fitting for zero
point, plate scale, and rotation resulted in a solution where the rms
residuals of individual stars are 0\farcs035 in right ascension and
0\farcs032 in declination; this is negligible compared to the X-ray
positional errors. The boresighted X-ray positions and error circles
were overlayed on this image, and photometry and proper motions for
the astrometric matches were extracted from the source catalog created
from the entire {\em HST} data set. We find seven new candidate
counterparts for six X-ray sources. The new matches are all likely
extra-galactic, given their proper motion or extended morphology.
From the fact that two of the seven {\em HST} candidate counterparts
are matched to {\em Chandra} sources with another candidate
counterpart, we estimate that an appreciable fraction of {\em HST}
matches could be random, on the order of 2/7 ($\approx$29\%) or even
more. As the astrometrically-calibrated image is not readily
available, we show the finding charts of these additional
identifications in Fig.~\ref{fig_hstid}. This brings the final tally
of the optical identification to 53 candidate counterparts for 48 of
the 57 X-ray sources inside $r_h$, and 14 candidate counterparts for
12 of the 29 sources between $r_h$ and 8\arcmin. The properties of the
candidate counterparts are summarized in Tables~\ref{tab_mem} and
\ref{tab_doubles}, and their locations in the optical color-magnitude
diagrams (CMDs) are shown in Figures~\ref{fig_cmd_mem} and
\ref{fig_cmd_nmem}. More information on the light-curve properties of
the optical variables (column 10 of Table~\ref{tab_mem}) can be found
by consulting the original references. Variables with names listed in
the format {\em nnnnn\_*} were first discovered by \cite{demaea07};
Table~1 in \cite{demaea07} gives an overview of the discovery papers
for variables with names starting with a 'V' or 'B', although updated
light curves can sometimes be found in more recent papers mentioned in
that table.

\begin{sidewaystable*}
\caption{Optical properties of candidate counterparts \label{tab_mem}}
\begin{center}
\begin{tabular}{rrccccccccclll}
\hline
\hline
(1) & (2) & (3)          & (4) & (5) & (6)   & (7)   & (8)          & (9)                           & (10)& (11)   & (12)             & (13)  & (14)    \\
CX  & OID & $d_{{\rm OX}}$ & $p_{\mu}$ & $V$ & $B-V$ & $V-I$ & \multicolumn{1}{c}{$L_{\rm X,u}$} & $\log (F_{\rm X}/F_{V})_{\rm u}$ & Var & $P$    & Optical spectrum & Class & Comments \\
    &     & (\arcsec)    &     &     &       &       & \multicolumn{1}{c}{(10$^{30}$ erg s$^{-1}$)}&                               &     & (days) &                  &       &         \\
\hline\\[-5pt]
\multicolumn{14}{c}{Proper-motion cluster members}\\[5pt]
\hline
 3~~ &  2893 & 0.23 & \ldots & 20.64 &$-$0.82 &  0.63 & 52.3 & $-$0.06 $\pm$ 0.04 & B8        & \ldots & \ldots &   CV & \ldots \\ 
 4~~ &  9315 & 0.09 &  m & 22.70 &  0.24 &  0.88 & 34.3 &    0.58 $\pm$ 0.04 & 06289\_9  & \ldots & emission lines   &  CV & \ldots \\ 
 9~~ & 10050 & 0.34 & 99 & 16.51 &  1.21 &  1.22 & 12.0 & $-$2.35 $\pm$ 0.08 & \ldots    & \ldots & early-K giant          &  RG & \ldots \\ 
15~~ &  6371 & 0.16 & 99 & 17.27 &  1.15 &  1.41 &  7.7 & $-$2.24 $\pm$ 0.09 & V9        &  3.187 & early/mid K, filled-in H$\alpha$ & SSG &  eclipsing binary \\ 
17~~ & 12652 & 0.26 & 99 & 17.86 &  0.27 &  0.26 &  6.4 & $-$2.1 $\pm$ 0.1 & B7        & \ldots & Balmer abs.~lines with ems.~cores &  CV &  \ldots \\ 
22~~ & 12695 & 0.75 & 99 & 17.80 &  0.94 &  1.02 &  5.4 & $-$2.2 $\pm$ 0.1 & V16       &  2.266 & \ldots &  AB & eclipsing or spotted binary \\ 
23~~ & 11111 & 0.21 & 98 & 17.13 &  0.99  &  1.07 &  4.6 & $-$2.5  $\pm$ 0.1  & V100      & 12.522    & late G/early K &  AB & spotted variable? \\ 
\ldots & \ldots & \ldots & \ldots & \ldots & \ldots  & \ldots & \ldots & \ldots  & \ldots      & or 23.947 & \ldots &  \ldots & \ldots \\ 
30~~ &  3626 & 0.31 & 99 & 17.96 &  1.15 &  1.29 &  4.8 & $-$2.8 $\pm$ 0.2 & V17       &  6.366 & late G, filled-in H$\alpha$? & SSG & \ldots \\ 
33~~ & 10611 & 1.30 & 99 & 19.57 &  1.04 &  1.20 &  3.3 & $-$1.7 $\pm$ 0.2 & V42       &  0.506 & \ldots &  AB & spotted binary \\ 
39~~ &  5883 & 0.38 & 99 & 17.20 &  0.89 &  0.94 &  2.6 & $-$2.8 $\pm$ 0.2 & V5        &  0.313 & mid G &  AB & contact binary \\ 
41~~ &  7011 & 0.23 & 99 & 18.27 &  1.06 &  1.12 &  2.3 & $-$2.4 $\pm$ 0.2 & V76       &  4.092 & late G, early K & SSG & \ldots \\ 
44~~ & 12390 & 0.62 & 99 & 17.89 &  0.93 &  1.01 &  2.2 & $-$2.5 $\pm$ 0.2 & V80       &  4.886 & \ldots &  AB & eclipsing binary \\ 
50~~ &  8001 & 0.28 &  m & 22.72 &  1.31 &  2.48 &  1.7 & $-$0.7 $\pm$ 0.2 & \ldots    & \ldots & late K/early M; H$\alpha$ emission &  AB & on binary main sequence \\ 
54~~ & 11278 & 0.39 & 96,m & 19.91 &  1.07 &  1.35 &  1.4 & $-$1.9 $\pm$ 0.3 & \ldots    & \ldots & K; filled-in H$\alpha$ &  AB? & on binary $V-I$ main seq. \\ 
57\tablenotemark{a} &  7397 & 0.20 & 77 & 19.27 &  1.15 &  1.32 &  1.1 & $-$2.3 $\pm$ 0.3 & V41       &  0.482 & \ldots &  AB & eclipsing or spotted binary \\ 
68~~ & 15561 & 0.92 & 99,m & 17.65 & \ldots&  1.53 & 12.7 & $-$1.9 $\pm$ 0.1 & 01431\_10 &  7.640 & \ldots & SSG & spotted binary \\ 
77~~ &   746 & 1.20 & 99,m & 17.96 &  1.35 &  1.39 &  4.5 & $-$2.2 $\pm$ 0.2 & V59       & 13.833 & mid K, weak H$\alpha$ emission & SSG & \ldots \\ 
79~~ & 13737 & 0.21 & 90 & 21.06 &  1.45  &  1.93 &  2.7 & $-$1.2  $\pm$ 0.2  & \ldots    & \ldots & \ldots &  AB & on binary main sequence \\ 
81~~ &  1185 & 0.79 &99 & 17.74 &  0.97  &  0.96 &  3.5 & $-$2.4  $\pm$ 0.2  & V7        &  0.488 & early/mid G &   AB & eclipsing binary \\ 
86~~ &  4773 & 1.93 & 99 & 17.55 &  1.04 &  1.03 &  3.5 & $-$2.5 $\pm$ 0.3 & V12       &  1.523 & mid G &  AB & eclipsing binary \\ 
\hline\\[-5pt]
\multicolumn{14}{c}{Sources with uncertain or no membership information}\\[5pt]
\hline\\[-5pt]

18 &  6697 & 0.13 &\ldots& 23.99 &$-$2.4: &  1.45 &  5.9 &    0.3  $\pm$ 0.1  & \ldots    & \ldots & \ldots &   FB & \ldots \\ 
19 & 12116 & 0.40 &\ldots& 23.71 &  0.21  &  1.30 &  5.9 &    0.2  $\pm$ 0.1  & \ldots    & \ldots & Balmer and He~II emission &  CV? & \ldots \\ 
25 &  5627 & 0.24 &\ldots& 24.09 &  0.4:  &  2.12 &  3.8 &    0.2  $\pm$ 0.1  & \ldots    & \ldots & \ldots &   FB & \ldots \\ 
28 & 10837 & 0.56 &\ldots& 22.71 & \ldots  &  0.38 &  3.8 & $-$0.4  $\pm$ 0.2  & \ldots    & \ldots & \ldots &   FB & \ldots \\ 
36 &  6751 & 0.90 &\ldots& 20.14 &  0.53  &  0.92 &  3.2 & $-$1.5  $\pm$ 0.2  & \ldots    & \ldots & \ldots&   FB & \ldots \\ 
49 &  5910 & 0.23 & 45   & 20.82 &  1.47  &  1.79 &  1.8 & $-$1.5  $\pm$ 0.2  & \ldots    & \ldots & mid K, H$\alpha$ emission &  AB? & on binary main sequence\\ 
56 &  4436 & 0.07 &\ldots& 20.83 &  1.29  &  1.69 &  1.1 & $-$1.7  $\pm$ 0.3  & 01724\_9  &  1.613 & \ldots &   AB & spotted variable \\ 
58 &  5913 & 0.35 &\ldots& 18.38 &  0.93  &  0.93 &  1.1 & $-$2.6  $\pm$ 0.3  & \ldots    & \ldots & \ldots &  AB? & on main sequence \\ 
65 & 14602 & 1.11 &\ldots& 23.05 &$-$0.4: &  0.54 &  9.0 &    0.1  $\pm$ 0.1  & 00670\_10 & \ldots & \ldots &   FB & long-term variable \\ 
72 & 13270 & 1.31 &\ldots& 22.9: &\ldots  &  0.5: &  5.4 & $-$0.1:            & \ldots    & \ldots & \ldots &   FB & \ldots \\ 
\hline
\multicolumn{14}{l}{Continued on next page}\\
\end{tabular}
\end{center}
\vspace{10cm}
\end{sidewaystable*}

\setcounter{table}{1}
\begin{sidewaystable*}
\caption{(Continued)}
\begin{center}
\begin{tabular}{rrccccccccclll}
\hline
\hline
(1) & (2) & (3)          & (4) & (5) & (6)   & (7)   & (8)          & (9)                           & (10)& (11)   & (12)             & (13)  & (14)    \\
CX  & OID & $d_{{\rm OX}}$ & $p_{\mu}$ & $V$ & $B-V$ & $V-I$ & \multicolumn{1}{c}{$L_{\rm X,u}$}  & $\log (F_{\rm X}/F_{V})_{\rm u}$ & Var & $P$    & Optical spectrum & Class & Comments \\
    &     & (\arcsec)    &     &     &       &       & \multicolumn{1}{c}{(10$^{30}$ erg s$^{-1}$)}&                               &     & (days) &                  &       &         \\
\hline\\[-5pt]
\multicolumn{14}{c}{Sources likely not associated with the cluster}\\[5pt]
\hline
 1~~ &  5361 & 0.19 & 59 & 20.36 &  0.65 &  0.76 &\ldots  & $-$0.09 $\pm$  0.03 & \ldots    & \ldots & AGN, $z\approx1.16$ &  EG  & \ldots \\ 
 2~~ &  3886 & 0.14 &  0    & 16.22 &  1.09 &  1.35 &\ldots  & $-$1.86 $\pm$  0.04 & V33       &  1.173 & mid K, H$\alpha$ emission &  AB  & eclipsing binary \\ 
 7~~ &  7878 & 0.17 &  0 & 15.39 &  1.21 &  1.49 &\ldots  & $-$2.68 $\pm$  0.06 & V19       &  \ldots & late-G, early-K foreground dwarf &  S   & irregular variable \\ 
 8~~ &  3472 & 0.17 &\ldots & 20.22 &  0.24 &  0.69 &\ldots  & $-$0.30 $\pm$  0.07 & \ldots    & \ldots & AGN, $z\approx2.36$ &  EG  & \ldots \\ 
12~~ & 12901 & 0.92 &  0 & 16.06 &  1.23  &  1.38 &  \ldots & $-$2.65 $\pm$ 0.09 & V66       & 50.498 & early-K foreground dwarf &   S  & long-period variable \\ 
13~~ &\ldots & 0.25 &\ldots & \ldots& \ldots&\ldots &\ldots  & $>$2.1              & \ldots    & \ldots & \ldots &  EG  & extended \\ 
21\tablenotemark{a} &\ldots & 0.08 & nm    & 23.96 & \ldots & 1.12&\ldots  &    0.2  $\pm$  0.1  & \ldots    & \ldots & \ldots & EG?  & proper motion consistent \\ 
\ldots &\ldots & \ldots & \ldots    & \ldots & \ldots & \ldots&\ldots  &   \ldots  & \ldots    & \ldots & \ldots & \ldots  & ~~with background galaxy \\ 
24~~ &  6271 & 0.29 & 27 & 16.32 &  0.89 &  1.01 &\ldots  & $-$2.9  $\pm$  0.1  & V1        &  0.268 & \ldots &  AB  & foreground contact binary \\ 
26~~ &\ldots & 0.22 &\ldots & 28.5  & \ldots& 1.5 &\ldots  &    1.9  $\pm$  0.2  & \ldots    & \ldots & \ldots &  EG  & extended \\ 
27\tablenotemark{a} & 11376 & 0.04 & 0 & 15.97 &  0.66 &  0.81 &\ldots  & $-$3.0  $\pm$  0.1  & V6        &  0.279 & \ldots &  AB  & foreground contact binary \\ 
29~~ &  7328 & 0.18 &  0    & 13.60 &  0.87 &  1.26 &\ldots  & $-$4.0  $\pm$  0.2  & \ldots    & \ldots & mid K &   S  & \ldots \\ 
31~~ & 11212 & 0.25 &\ldots & 22.27 &  0.51 &  0.75 &\ldots  & $-$0.6  $\pm$  0.2  & \ldots    & \ldots & AGN, $z\approx1.53$ &  EG  & \ldots \\ 
34\tablenotemark{a} &\ldots & 0.33 &\ldots &\ldots &\ldots &\ldots &\ldots  &  $>$1.6             & \ldots    & \ldots & \ldots &  EG  & extended \\ 
35~~ &  4571 & 0.28 &  0    & 14.62 &  1.38 &  1.35 &\ldots  & $-$3.7  $\pm$  0.2  & \ldots    & \ldots & mid-K giant &   S  & \ldots \\ 
38~~ &  7534 & 0.12 & 11,nm    & 21.04 &  1.57 &  2.29 &\ldots  & $-$1.2  $\pm$  0.2  & \ldots    & \ldots & \ldots &   S  & \ldots \\ 
42~~ &\ldots & 0.15 &\ldots & 26.6  &\ldots & 0.7 &\ldots  &    1.0  $\pm$  0.3  & \ldots    & \ldots & \ldots &  EG  & extended \\ 
43~~ &  3629 & 0.78 & 98 & 22.11 &  1.75  &  2.94 &  \ldots & $-$0.6  $\pm$ 0.2  & \ldots    & \ldots & \ldots &   S & too red to be a member \\ 
45~~ &  4353 & 0.37 &  0    & 18.98 &  1.45 &  1.57 &\ldots  & $-$1.7  $\pm$  0.2  & \ldots    & \ldots & $z\approx0.092$ &  EG  & extended \\ 
46~~ & 10088 & 0.21 &  0 & 11.54 &  1.10 &  1.10 &\ldots  & $-$5.1  $\pm$  0.2  & \ldots    & \ldots & mid-G foreground giant &   S  & \ldots \\ 
47~~ & 10967 & 0.32 & nm    & 23.88 &$-$0.2 &  2.09 &\ldots  & $-$0.2  $\pm$  0.2  & \ldots    & \ldots & \ldots &  EG  & extended \\ 
48~~ &  2426 & 0.66 &  0 & 19.53 &  1.04  &  1.15 &  \ldots & $-$1.8  $\pm$ 0.2  & V11       &  0.883 & early G &   AB & eclipsing binary \\ 
51~~ & 12037 & 0.41 &  0    & 14.01 &  0.71 &  0.74 &\ldots  & $-$4.2  $\pm$  0.2  & \ldots    & \ldots & late-F/early-G dwarf &   S  & \ldots \\ 
52~~ &\ldots & 0.18 &\ldots & 27.68 &\ldots & 2.32&\ldots  &    1.2  $\pm$  0.3  & \ldots    & \ldots & \ldots &  EG  & extended \\ 
53~~ &  8705 & 0.20 & nm    & 23.7  &  0.3  &  1.3  &\ldots  & $-$0.4  $\pm$  0.3  & \ldots    & \ldots & \ldots &  EG  & extended \\ 
55~~ &\ldots & 0.80 &\ldots & 18.76\tablenotemark{b}  & \ldots  &  \ldots  &\ldots  & $-$2.4$\pm$ 0.3  & \ldots    & \ldots & \ldots &  EG?  & extended \\ 
76~~ & 14833 & 2.37 & 0 & 16.48 &  1.00  &  1.05 & \ldots & $-$2.9  $\pm$ 0.2  & V54       &  8.314 & mid-G, early-K foreground dwarf &   AB & spotted variable? \\ 
78\tablenotemark{a} &   553 & 2.58 & 0 & 16.85 &  1.03 &  1.10 &\ldots  & $-$2.7  $\pm$  0.2  & V23       &  0.272 & \ldots  &  AB  & foreground contact binary \\ 
80~~ &  1031 & 2.05 & 0 & 20.36 &  0.85  &  0.94 & \ldots & $-$1.3  $\pm$ 0.2  & \ldots    & \ldots & \ldots &   S & \ldots \\ 
82\tablenotemark{a} & 13404 & 0.03 & 0 & 14.47 &  0.87 &  0.89 &\ldots  & $-$3.9  $\pm$  0.2  & \ldots    & \ldots & early-K foreground dwarf &   S  & \ldots \\ 
84\tablenotemark{a} & 15434 & 1.49 & 0 & 21.2: &\ldots  &  2.9: &  \ldots & $-$1.2  $\pm$ 0.3  & 01225\_10 & \ldots & \ldots &   S &  long-term variable \\ 
\hline
\end{tabular}
\end{center}

$^a$This source has another candidate counterpart, see
Table~\ref{tab_doubles}.\\ $^b$The counterpart is not included in the
S03 catalog. The listed value is the magnitude in the $g\prime$ band
\citep{platea11}.\\

\tablecomments{Columns: 1) source number; 2) number of candidate
  optical counterpart in the S03 catalog, sources without a number
  were found in the {\em HST}/ACS images or (for CX\,55 only) by
  visual inspection of the S03 cluster image; 3) angular separation
  between the X-ray and optical source {\em after correcting for
    boresight}; 4) probability for cluster membership, where numeric
  values are based on the proper-motion studies of \cite{platea11}
  (the case of CX\,3 is described in more detail in
  Sect.~\ref{sec_cv}) and K.\,Cudworth et al. (private communication),
  and the labels ``m'' (``members'') and ``nm'' (``(non)-members'')
  are based on the {\em HST}/ACS study described in \cite{bediea06};
  5--7) optical magnitudes and colors from S03, except for the
  (unnumbered) {\em HST} counterparts where we list the F606W
  magnitude and the F606W~$-$~F814W color if available; 8) unabsorbed
  X-ray luminosity (0.3--7 keV) in units of $10^{30}$ erg s$^{-1}$
  assuming a distance of 4.1 kpc and the fluxes in
  Table~\ref{tab_src}; 9) ratio of X-ray (0.3--7 keV) to $V$ band
  fluxes after correcting for extinction; 10) name of the associated
  optical variable; 11) period of variability; 12) spectroscopic
  properties of the candidate optical counterpart; 13) source class:
  CV and CV? = (candidate) cataclysmic variable, AB and AB? = likely
  or candidate active binary, SSG = sub-subgiant, RG = red giant, S =
  likely foreground star not associated with the cluster without any
  indication of binarity, EG = extra-galactic source, FB = Faint
  object to the Blue of the main sequence ; 14) comments. Information
  on optical variability was gathered from the following references:
  \cite{kaluruci93}, \cite{ruciea96},
  \cite{mochea02,mochea03,mochea05}, \cite{brunea03}, \cite{hartea05},
  \cite{demaea07}, \cite{brogea11}}
\vspace{-10cm}
\end{sidewaystable*}

\subsubsection{Sources with multiple counterparts} \label{sec_doubles}

For seven X-ray sources we find two candidate counterparts. In four
cases---CX\,27, CX\,57, CX\,78, CX\,84---one of the two is an optical
variable, and given the low probability for a chance coincidence we
consider the variables to be the true matches. The variables all lie
closer to the X-ray source than the alternative match. We also
tentatively list the closest match to the other three X-ray sources as
the likely counterpart in Table~\ref{tab_mem}, but more information is
needed to firmly establish which, if any, of the two is the true
match. For completeness we list the properties of the alternative
identifications in Table~\ref{tab_doubles}. CX\,34 is likely an
extra-galactic source, as both {\em HST} counterparts are
extended. The closest match to CX\,21 is a faint point source that is
unrelated to the cluster; its {\em HST} proper motion is typical for a
background galaxy. The other match is a proper-motion member on the
main sequence (see also Fig.~\ref{fig_hstid}). CX\,82 matches with two
bright ($V<15$) stars. The nearest is a proper-motion non-member and a
likely foreground K dwarf.

\begin{table}
\caption{Alternative optical counterparts\label{tab_doubles}}
\begin{center}
\begin{tabular}{crccccl}
\hline
\hline
CX & \multicolumn{1}{c}{OID} & $d_{\rm OX}$ & $V$ & $B-V$ & $V-I$ & comment\\ 
   &     & (\arcsec)   &    &       &       &        \\
\hline
21 &     11098  &   0.78 &    18.48  &  0.93   &    0.97 &  $p_{\mu}=90$\\
27 &     11365  &   0.68 &    16.58  &  0.76   &    0.94 &  $p_{\mu}=0$, F star\\
34 &    \ldots  &   0.68 &    \ldots & \ldots  &   \ldots&  extended\\
57 &      7395  &   0.76 &    21.02  &  1.47   &    1.96 &  non-member, K star\\
78 &       569  &   3.12 &    16.50  &  0.95   &    1.03 &  $p_{\mu}=0$\\
82 &     13395  &   0.70 &    14.66  &  0.96   &    0.98 &  \ldots\\
84 &     15442  &   3.01 &    20.64  & \ldots  &    1.7  &  \ldots\\
\hline
\end{tabular}
\tablecomments{See Table~\ref{tab_mem} for the closest astrometric
  optical match to the X-ray source, and for the meaning of the
  columns.}
\end{center}
\end{table}

\subsection{Source classification}

\begin{figure*}
\centerline{
\\
\includegraphics[width=8.5cm]{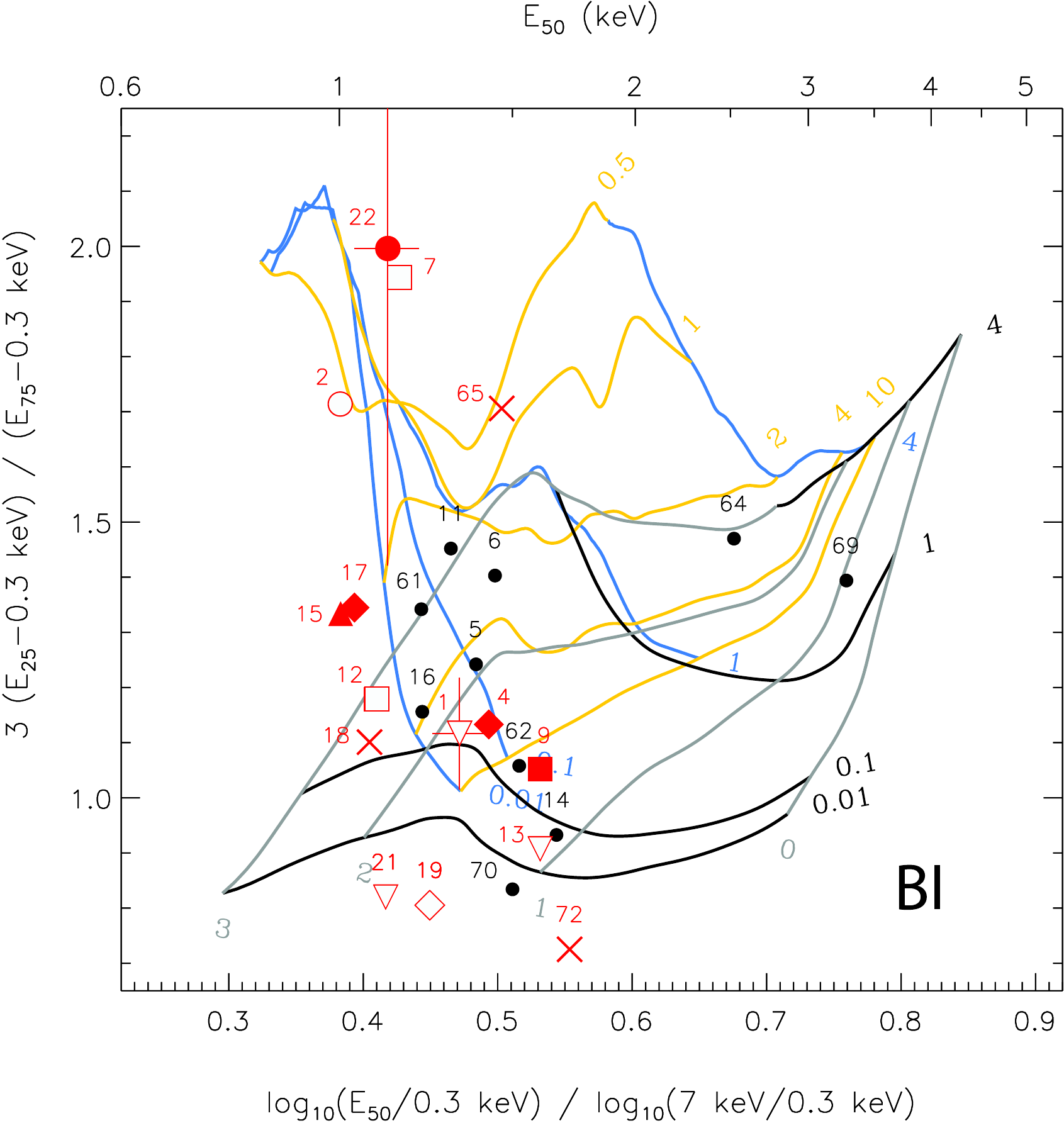}
\hspace{0.3cm}
\includegraphics[width=7.93cm]{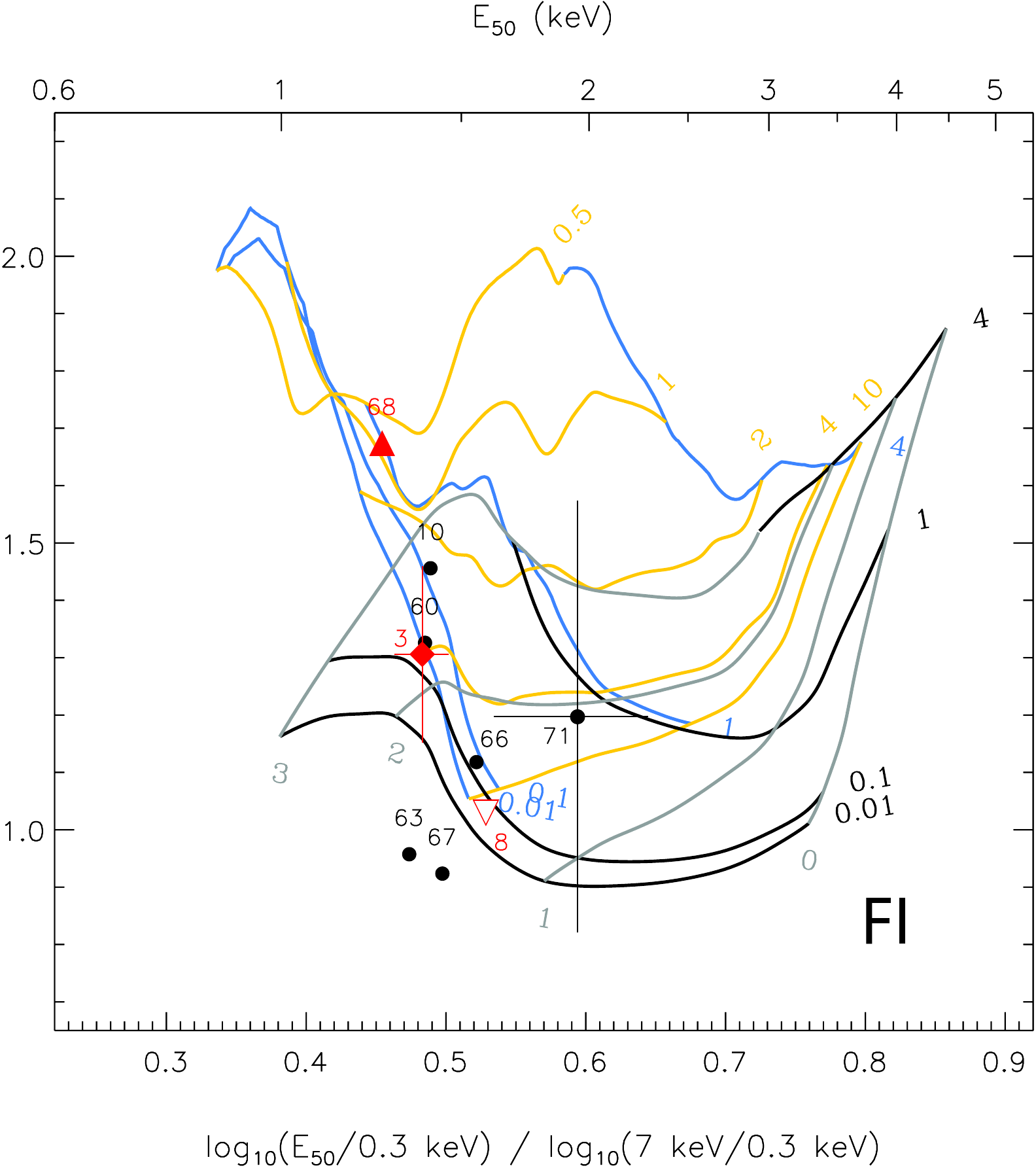}
}
\caption{Quantile color-color diagrams with model grids representing a
  Mekal plasma (blue/yellow) and a power-law spectrum
  (black/grey). For a certain choice of model, the plasma temperature
  or spectral index, and the column density can be estimated from the
  location of a source inside the grid. The median energy $E_{50}$ can
  be read off from the top x-axis. Symbols are as in
  Figures~\ref{fig_cmd_mem} and \ref{fig_cmd_nmem}, while black filled
  circles mark sources without optical counterparts. Due to their
  different energy responses, sources on the background-illuminated
  (BI) S\,3 chip and foreground-illuminated (FI) S\,2 and S\,4 chips
  are shown in separate panels. We include sources with more than 15
  net counts (0.3--7 keV); error bars are shown only for the sources
  with the highest and lowest number of counts. \label{fig_qccd}}
\end{figure*}

Many factors contribute to the classification of the X-ray sources. An
extended morphology of the candidate optical counterpart separates
extra-galactic from galactic sources. Based on optical spectra one can
immediately distinguish between AGN, accreting binaries with
substantial accretion disks, and ordinary stars. Period and shape of
the optical light curve constrain the nature and period of a close
binary. Proper-motion information, and to some extent optical colors,
can establish cluster membership and puts limits on the distance. To
distinguish between cluster members and non-members we mainly rely on
the proper-motion study by \cite{platea11}, which includes 41 of our
candidate counterparts.  For nine additional candidate counterparts
membership could be established from the {\em HST} study by Bedin et
al.~(2006, see Sect.~\ref{sec_hst}) and from the results of the
proper-motion study by K.~Cudworth (private communication).

We use the energy quantiles to constrain the underlying X-ray spectrum
for sources with more than 15 net counts (0.3--7 keV).  For fainter
sources the errors on the quantiles are too large to meaningfully say
anything about their spectra.  Quantile color-color diagrams (QCCDs)
are shown in Fig.~\ref{fig_qccd}. Each panel shows two types of model
grids, which indicate the expected locations of a source in the QCCD
for emission from a Mekal plasma (blue/yellow) and for a power-law
spectrum (black/gray). The former is appropriate for the emission from
hot coronae around active stars for which the plasma temperature $kT$
can reach up to several keV. CVs can be anywhere between soft or very
hard; the spectrum of a typical dwarf nova has $kT\approx2-10$ keV
\citep[e.g.][]{byckea2010}. Power-law spectra are appropriate for the
harder emission from AGN ($\Gamma=1-2$). Grids are computed with {\tt
  Sherpa} using the energy response of the source closest to the
aimpoint. Variations in the shape of the grids as function of chip
position are typically smaller than the errors on the measurements,
and are ignored.

The ratio of the fluxes in the X-ray and optical ($V$) band, or limits
thereon, can help to classify a source, even in the absence of an
optical match down to the limit of the observations. We compute this
ratio as follows: $\log (F_{\rm X}/F_{\rm opt})_{\rm u} = \log (F_{\rm
  x})_{\rm u} + (V-A_V)/2.5 + 5.44$. The zeropoint for the $V$-band
flux is taken from \cite{bessea}. The total extinction $A_V$ was
assumed to be equal to the cluster value, which underestimates
(overestimates) the flux ratio for foreground (background) objects.
Typically, stars and active binaries have $\log (F_{\rm X}/F_V)_{\rm
  u} \lesssim -1$, while AGN, CVs or other accreting binaries with
unevolved late-type companions, and very active late-type dwarfs can
have higher ratios \citep[e.g.][]{stocea91}.

\section{Results} \label{sec_results}

\subsection{Cluster members versus non-members} \label{sec_membership}

Twenty candidate counterparts are proper motion members. Of these, all
but one (CX\,9) show signs of binarity. On this basis we consider at
least 19 of them as the true counterparts to the {\em Chandra}
sources. We derive the binary status and type of binary from the
optical spectra, the optical light curves, or a location in the CMD on
the binary main sequence, indicating that an unresolved multiple
system is responsible for the detected light. The cluster binaries are
a mix of CVs, ABs, and binaries below the sub-giant branch, and are
discussed further in Sects.~\ref{sec_cv}, \ref{sec_ab}, and
\ref{sec_ssg}, respectively. CX\,9 is discussed in
Sect.~\ref{sec_single}. Cluster members are marked with filled symbols
in the CMDs of Fig.~\ref{fig_cmd_mem}.

A total of 30 sources are not, or likely not, associated with the
cluster. These are stars and binaries with proper motions that clearly
set them apart from the members, background galaxies, and stars whose
very red colors classify them as unlikely members. They are shown in
Fig.~\ref{fig_cmd_nmem} and briefly discussed in
Sects.~\ref{sec_single}, \ref{sec_galaxies}, and \ref{sec_uncertain}.

For ten candidate counterparts membership information is lacking or
inconclusive. These sources include (candidate) ABs, a new candidate
CV, and faint optical sources that lie to the blue of the main
sequence; the latter are discussed in Sect.~\ref{sec_uncertain}. Open
symbols and crosses mark them in Fig.~\ref{fig_cmd_mem}.

In the following we discuss the sources by type of X-ray
emitter. Unless stated otherwise, X-ray luminosities quoted in the
text refer to the 0.3--7 keV band.

\subsection{Cataclysmic variables} \label{sec_cv}

We have detected three CVs that belong to the cluster and discovered
one CV candidate without membership information. For all four, the
X-ray colors and luminosities are in the expected range for CVs. While
the X-ray--to-optical flux ratios in Table~\ref{tab_mem} are also
typical for CVs, their values can be misleading: the X-ray and optical
data are not contemporaneous, while CVs may show large variations in
brightness on a time scale of weeks to months.

\begin{figure*}
\centerline{\includegraphics[width=16cm]{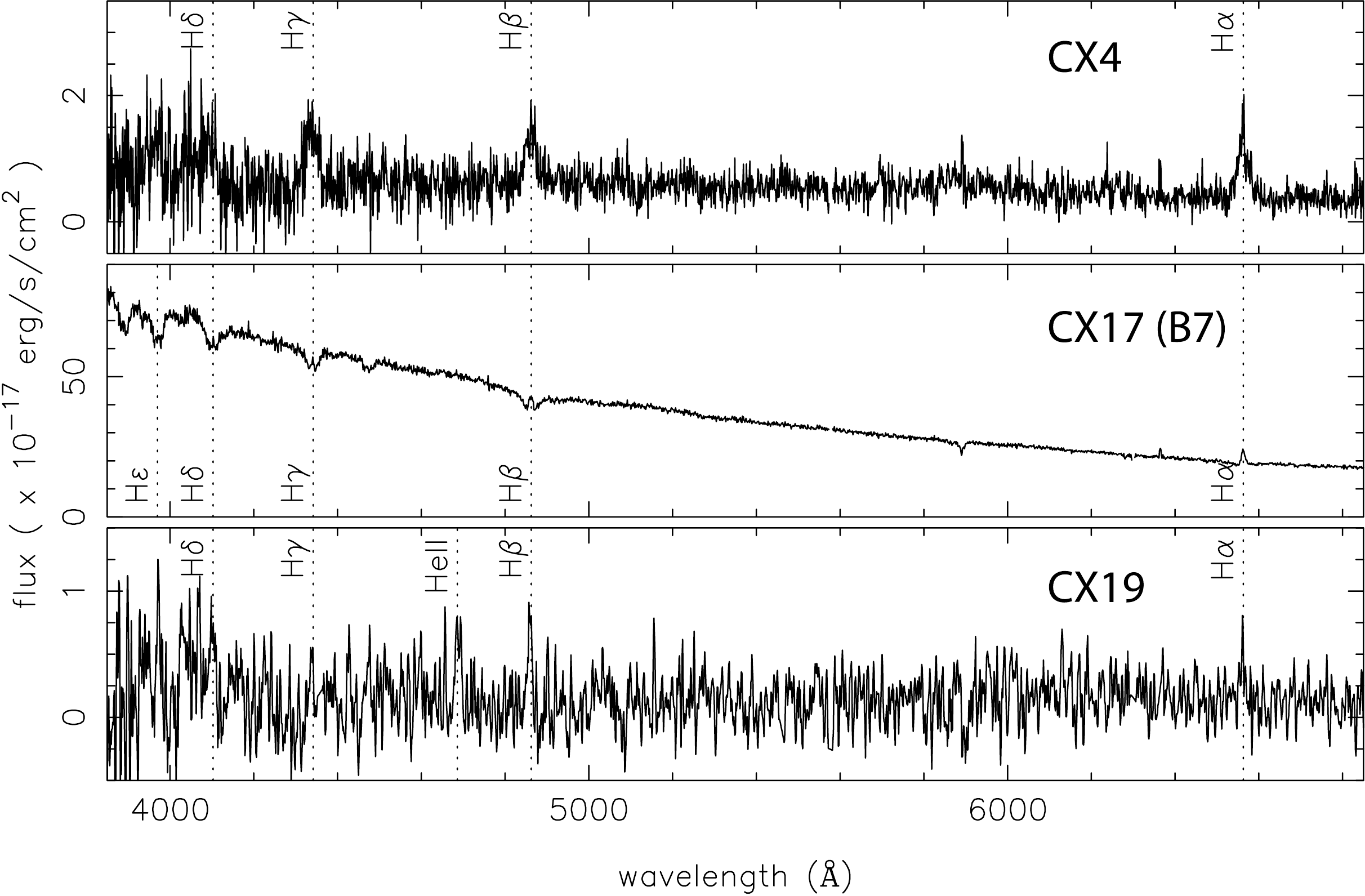}}
\caption{Hectospec spectra of two CVs and one candidate CV. Top: our
  optical spectrum of CX\,4 = 06289\_9 confirms this optical variable
  and cluster member is a CV, as was already suggested by
  \cite{demaea07}. Middle: the spectrum of the known member CV CX\,17
  = B7 shows narrow emission lines superposed on a continuum with
  broad Balmer absorption features, indicating a high accretion
  rate. Bottom: the new candidate CV CX\,19. To reduce noise, its
  spectrum has been smoothed with 3 pixels ($\Delta \lambda \approx
  3.6$ \AA). Regions where imperfect sky subtraction left large
  residuals (around 5577\AA~and 6300\AA) have been
  removed.  \label{fig_cvspec}}
\end{figure*}

CX\,4 is matched to the faint cluster member and optical variable
06289\_9 that lies to the blue of the cluster main sequence. Based on
its optical colors and the detection of an outburst-like event,
\cite{demaea07} already suggested that this star is a dwarf nova. To
our knowledge the Hectospec spectrum in Fig.~\ref{fig_cvspec}, which
shows the Balmer lines clearly in emission, is the first spectroscopic
confirmation of its CV nature. The X-ray quantiles suggest that the
X-ray emission arises in a $kT\approx4-10$ keV plasma
(Fig.~\ref{fig_qccd}). Thus $L_{\rm X}$ in Table~\ref{tab_mem}, which
was computed for the assumption of a 2-keV plasma, can be up to
$\sim$40\% too low (see Sect.~\ref{sec_xspec}).

The new candidate CV is the X-ray variable CX\,19, which is matched to
a faint blue object with a Hectospec spectrum that shows He\,II 4686
\AA~and H$\beta$ emission lines (Fig.~\ref{fig_cvspec}), and hints of
H$\alpha$ and H$\delta$ in emission. The X-ray luminosity ($L_{{\rm
    X,u}} \approx 6 \times 10^{30}$ erg s$^{-1}$), the hard spectrum
as suggested by its quantile values (Fig.~\ref{fig_qccd}), and the
high $(F_{\rm X}/F_{\rm V})$ ratio, are all consistent with a
classification as CV. He\,II 4686 \AA~is of comparable strength as
H$\beta$. This is seen in high mass-transfer rate systems (nova-like
CVs), and in CVs containing white dwarfs with much stronger magnetic
fields (i.e.~$\gtrsim$1 MG) than in dwarf novae systems. If we assume
cluster membership we find an absolute magnitude $M_V\approx10.2$,
which favors the explanation as a magnetic CV.

CX\,3 and CX\,17 are the known, spectroscopically-confirmed CVs B8 and
B7 \citep{kaluea97}. With $L_{{\rm X,u}} \approx 5.2 \times 10^{31}$
erg s$^{-1}$, B8 is the brightest X-ray source in the cluster. The
actual value of $L_X$ is likely higher, as the X-ray quantiles suggest
a plasma temperature ($\sim$4 keV) that is higher than our nominal
value. B8 has been classified as a SU\,UMa dwarf nova based on the
detection of several outbursts of 1--2 mag in amplitude, and a recent
superoutburst ($\sim$3 mag) in the Kepler light curve
\citep{mochea03,garnea11}.  We include B8 among the cluster
members. \cite{platea11} estimated a 10\% membership probability from
proper motion, but closer inspection of the data reveals that a few
detections of this star were contaminated by a fainter close
neighbor. When the affected data points are not considered, the proper
motion of B8 is consistent with cluster membership.

CX\,17 or B7 is a solid cluster member. The X-ray quantiles indicate
that $kT \approx 2$ keV. Most reports in the literature find B7 to be
relatively bright with $V\approx18$, which suggests a high
mass-transfer rate; occasional small (0.5--1 mag) outbursts, and a
drop in brightness of $\sim$3 mag in $V$ have also been seen.
\cite{mochea03} propose that B7 is a nova-like variable of the VY Scl
type, or perhaps a Z\,Cam dwarf nova mostly seen during periods of
standstill between minimum and maximum brightness. Our Hectospec
spectrum is qualitatively similar to the spectrum in \cite{kaluea97},
and likely taken when the mass-transfer rate was high and the disk
optically thick: the spectrum shows broad Balmer absorption lines with
narrow emission cores, with the emission component dominant in
H$\alpha$ (Fig.~\ref{fig_cvspec}).

The X-ray--to--optical flux ratios and blue optical colors of CX\,18,
CX\,25, CX\,28, CX\,36, CX\,65, and CX\,72 resemble those of CVs, and
these sources are prime targets for optical follow-up spectroscopy.

\subsection{Active binaries} \label{sec_ab}

We classify a total of twenty sources as likely or candidate
ABs. Thanks to the extensive photometric variability studies of
NGC\,6791, the orbital periods of many of these can be inferred from
the light curves, either through eclipses, ellipsoidal variations, or
spot activity. Among the cluster members we find eleven ABs, including
eight with photometric periods (Table~\ref{tab_mem}). Most have colors
that place them near the main sequence or, in the case of the W\,UMa
contact binary CX\,39, right at the turnoff. For CX\,86, the 1-sigma
errors on $B-V$ and $V-I$ are such that it could lie anywhere from
just below the turnoff to the base of the red-giant branch. CX\,23
lies securely above the sub-giant branch, and we discuss it separately
below. NGC\,6791 is at least 8 Gyr old; by comparison to the orbital
period versus eccentricity distribution of solar-type spectroscopic
binaries in the 6.5-Gyr old open cluster NGC\,188 it is expected that,
for main-sequence binaries, orbits up to at least 15 days have been
circularized at the age of NGC\,6791 \citep{mathea04}. Therefore,
since tidal synchronization operates on a shorter time scale than
tidal circularization \citep{hut,zahn89}, main-sequence binaries with
periods below 15 days are also expected to be tidally locked. All of
our main-sequence ABs with measured photometric periods have periods
shorter than 4.9 d; in Sect.~\ref{disc_ab} we discuss how this could
be due to the period versus X-ray luminosity relation of ABs and our
detection sensitivity.  CX\,50, CX\,54, and CX\,79 are not identified
with known variables, but they lie on the binary main sequence or have
H$\alpha$ in emission in their spectra, which is a signature of
enhanced magnetic activity. We note that in the $V$ versus $B-V$ CMD,
CX\,54 lies slightly to the blue of the main sequence. This is not
expected for the combined light of two main-sequence stars, so its
classification as AB is less secure. Perhaps this star is a non-member
after all, with a proper motion that is similar to that of the cluster
stars. Radial-velocity measurements can provide more clarity.

Stars that have evolved past the turnoff, i.e.~have evolved to larger
radii, can in principle circularize wider orbits than main-sequence
stars within a given time span as the circularization time scale is
very sensitive to radius. For the sub-giant AB CX\,23 the eccentricity
is unknown, and estimates of the binary period are only known from
photometric variability. \cite{brunea03} quote a period of $\sim$12.5
d, while \cite{mochea05} find nearly twice that value, $\sim$23.9
d. From the $V$ and $B-V$ measurements of the optical counterpart, we
estimate that the radius of the primary is at most $\sim$2.8
$R_{\odot}$ \citep{flow96}; as both components of the binary likely
contribute to the measured optical light, the actual radius of the
primary is smaller. If we now look at the diagnostic diagram in
Verbunt \& Phinney (1995, Fig.~5a; appropriate for a primary mass of
1.25 $M_{\odot}$, i.e.~somewhat larger than the turnoff mass in
NGC\,6791), it is clear that, to first approximation, the primary in
CX\,23 is too small to have circularized a 23.9-d orbit. This means
that in order to explain the X-rays, the rotation of the primary would
have to be synchronized to the orbital motion around periastron,
assuming the binary was not circular to begin with. However, this
leaves the photometric period, which has been tentatively attributed
to spot activity and thus reflects the primary's rotation period,
unexplained. The shorter period found by \cite{brunea03} is more
consistent with the detection of X-ray activity, as a 12.5 d orbit is
short enough for tidal synchronization and circularization to have
been achieved. The level of X-ray activity is still unusually high for
such a period (Sect.~\ref{disc_ab}). With radial-velocity measurements
the exact period and eccentricity of the orbit can be verified.

\nocite{verbphin}

Three candidate counterparts without or with ambiguous membership
information are classified as (candidate) ABs near the main
sequence. CX\,56 is a 1.6-d spotted variable. If CX\,49 is a member,
it is almost certainly an AB owing to its location on the binary main
sequence and H$\alpha$ emission. CX\,58, for which we do not have an
optical spectrum, lies on the cluster main sequence but could still be
an AB in the cluster if the companion is faint.

Six proper-motion non-members are classified as ABs. These include the
W\,UMa contact binaries CX\,24, CX\,27, and CX\,78. The distance
modulus inferred from the relation between period, color, and absolute
magnitude of contact binaries also place them in front of the cluster
\citep{mochea03}.

\subsection{Sub-subgiants} \label{sec_ssg}

Five sources are matched to proper-motion cluster members ($p_m=99$\%)
that stand out due to their optical photometry, which places them
below or to the red of the sub-giant branch: CX\,15, CX\,30, CX\,41,
CX\,68, and CX\,77. Stars with similar properties have been labeled
``sub-subgiants'' or ``red stragglers'' owing to their unusual
location in the CMD, which cannot be explained by the combined light
of two ordinary cluster members. About two dozen sub-subgiants are
known in open and globular clusters. Various mechanisms (past or
ongoing mass transfer, dynamical encounters) have been suggested to
explain their photometry but most systems still defy explanation.

All five X-ray detected sub-subgiants in NGC\,6791 are optical
variables with periods between 3.2 and 13.8 d, and their puzzling
photometry has been pointed out before \citep{kalu03,platea11}. CX\,15
is matched to the eclipsing binary V9. Its striking light curve shows
smooth, asymmetric brightness modulations that migrate in phase on a
time scale of months; these have been attributed to large star spots,
and indicate a high level of activity \citep[see for
  example][]{mochea05}. The other four systems show a single maximum
and minimum in the light curve when folded on the periods cited in
Table~\ref{tab_mem}, and have been classified as plausible rotational
spotted variables \citep{mochea02,kalu03,demaea07}; unambiguous proof
of their binarity and orbital periods requires radial-velocity
follow-up. While the photometry of CX\,41 can be accounted for by the
light of three stars---in a bound multiple system or aligned by
chance---the red colors of the other four require at least one
anomalous star like an underluminous (sub)giant or an overluminous
main sequence star. Their detection as X-ray sources indicates ongoing
interaction between companions in a close binary. The weak H$\alpha$
emission in the optical spectra, and the relatively soft X-ray colors
are more in line with coronal activity than with an accretion origin
of the X-rays.  Two more sub-subgiants in NGC\,6791 were pointed out
by \cite{platea11}. Star 13753 from S03 falls outside the {\em
  Chandra} field. Star 83 is not detected but it lies almost
9\arcmin~from the observation aimpoint so was observed with low
sensitivity.

\subsection{Stars without indications of binarity} \label{sec_single}

CX\,9 matches with an early-K giant and cluster member. Our optical
spectrum shows no indications of activity, nor is this star known to
be an optical variable. Its X-ray colors are unusually hard for a
coronal source (Fig.~\ref{fig_qccd}). This giant is an interesting
target for follow-up studies to investigate if it is a symbiotic-like
interacting binary, in which the X-rays are powered by
accretion. Alternatively, a faint quasar could be hidden under the
point-spread function of this bright star, so a possible explanation
is that the giant is a spurious match.

Proper motions exclude cluster membership for the candidate
counterparts to CX\,7, CX\,12, CX\,29, CX\,35, CX\,38, CX\,46, CX\,51,
CX\,80, CX\,82, and CX\,84. These sources are plausible late-type
field stars. The proposed counterparts for CX\,7, CX\,12, and CX\,84
are reported to be long-term or irregular variables, but the light
curves do not point at binarity; the remaining stars are not known to
be variables. Our FAST spectra indicate that CX\,35 and CX\,46 are
(sub)giants. Our assessment of luminosity class is mainly based on the
absence or presence of the CN 4216 \AA~bandhead
\citep{graycorb09}. The $\log (F_{\rm X}/F_V)_{\rm u}$ values, and the
soft X-ray colors of CX\,7 and CX\,12 agree with a coronal origin of
the X-rays.

The colors of the faint source CX\,43 place it about 0.8 to the red of
the lower main sequence and make it an unlikely cluster member. On the
other hand, its proper motion is consistent with that of cluster
stars. A radial-velocity based membership constraint can provide a
more definite answer. Based on the red colors and high $\log (F_{\rm
  X}/F_V)_{\rm u}$ ratio, we suggest this source is a vey active
late-type star.

\subsection{Background galaxies} \label{sec_galaxies}

We classify twelve sources as extra-galactic on various grounds. The
optical spectra of CX\,1, CX\,8, and CX\,31 have broad emission lines
typical for AGN, and we estimate the redshift $z$ (see
Table~\ref{tab_mem}) by comparison with the quasar composite spectrum
from \cite{vandea01}. The counterpart to CX\,45 has a redshifted
stellar absorption-line spectrum and clearly is extended, even in the
ground-based image from S03. The proper motion of the faint unresolved
counterpart of CX\,21 is consistent with those of background galaxies
in the field \citep{bediea06}, but we note this source has a bright
alternative counterpart on the edge of the error circle that is a
cluster member (Sect.~\ref{sec_doubles}). CX\,13, CX\,26, CX\,34,
CX\,42, CX\,47, CX\,52, and CX\,53 look extended in the ACS images.

Apart from CX\,13 and CX\,34, which are too faint to get reliable
optical photometry for, the extra-galactic sources lie away from the
cluster main sequence and most have $\log (F_{\rm X}/F_V)_{\rm u}
\gtrsim -0.6$. We note that the value of $N_{\rm H}$ used to calculate
this ratio can be underestimated (and the flux ratio therefore
overestimated) since extra-galactic sources can have significant
intrinsic absorption.

\subsection{Uncertain classifications and unidentified sources} \label{sec_uncertain}

Six sources are matched to faint, blue objects without membership
information: CX\,18, CX\,25, CX\,28, CX\,36, CX\,65, and CX\,72. As
mentioned in Sect.~\ref{sec_cv}, their optical colors and
X-ray--to--optical flux ratios are compatible with those of CVs or
other accreting binaries with low-mass companions, but also with
AGN. Follow-up spectroscopy is needed to further investigate their
classification.

A total of 27 sources remain without candidate optical
counterparts. For these sources we estimate a lower limit on the
X-ray--to--optical flux ratio by assuming $V\gtrsim24$ (i.e.~the limit
of the S03 catalog), which gives $\log (F_{\rm X}/F_V)_{\rm u} \gtrsim
-0.9$.  This excludes a stellar coronal origin unless they are very
active (flaring) K or M dwarfs, but leaves open the possibility of
accreting compact binaries. An extra-galactic nature is the most
likely option based on the expected number of AGNs.  We compute this
number using the cumulative source density versus flux ($\log N - \log
S$) curves from \cite{kimea07}, which are derived from {\em Chandra}
pointings at high Galactic latitudes. We assume a 3-count detection
limit for the region within 3\arcmin~from the aimpoint, a 5-count
limit for larger offsets, and an intrinsic power-law spectrum with
photon index $\Gamma=1.7$ that is absorbed by the integrated Galactic
column density in the direction of the cluster ($9 \times 10^{20}$
cm$^{-2}$). This predicts $\sim$56 AGN. For comparison, we detect 12
galaxies. This suggests that most of the unclassified and unidentified
sources (33 in total) are AGN.

\section{Discussion} \label{sec_disc}

\subsection{Cataclysmic variables in NGC\,6791} \label{disc_cv}

NGC\,6791 is the open cluster where most CVs have been found so
far. This may not be surprising because, of the open clusters that are
old enough to find close binaries in a phase where they can be
observed as CVs, NGC\,6791 is the most massive one. We check if
primordial binaries can account for all observed systems, which
comprise the three members (the dwarf novae B8 and CX\,4, and the
nova-like variable or dwarf nova B7) and the new (possibly magnetic)
CV candidate CX\,19. Recent observational estimates of the CV number
density in the local field differ by a factor of a few but are
consistent within the (appreciable) errors: $4^{+6}_{-2} \times
10^{-6}$ pc$^{-3}$ \citep[][non-magnetic CVs only]{pretea12},
$0.9^{+1.5}_{-0.5} \times 10^{-5}$ pc$^{-3}$ \citep{rogeea08},
$\sim$$1 \times 10^{-5}$ pc$^{-3}$ \citep{grinhongea05}. Adopting the
local stellar mass density of 0.045 $M_{\odot}$ pc$^{-3}$ from
\cite{robiea03}, and scaling the resulting number of field CVs per
unit mass to the cluster mass ($\sim$5000--7000 $M_{\odot}$, see
Sect.~\ref{disc_oc}), we expect 0.2--3.7 CVs in the cluster. Therefore
the observed number of CVs in NGC\,6791 is consistent with a
primordial population and does not require any dynamical formation or
destruction processes.

It is possible that there are more CVs in NGC\,6791 that we missed
because they are too faint in either X-rays or the
optical. \cite{byckea2010} studied a sample of twelve dwarf novae
within $\sim$200 pc with parallax distances. Only two (GW\,Lib and
WZ\,Sge) have X-ray luminosities very close or below our detection
limit, which is $\sim$$1\times10^{30}$ erg s$^{-1}$ (0.3--7 keV) for
the typically hard spectra of CVs.  To our knowledge, there are no
similar, published, studies of the X-ray luminosity function for
magnetic CVs. Systems with moderate magnetic-field strengths (between
1--10 MG, the so-called intermediate polars) are thought to be
generally brighter in X-rays than dwarf novae due to the higher
mass-accretion rates; see e.g.~Heinke et al.\,(2008). Polars
($B\gtrsim10$ MG) on the other hand can be faint when in the low state
with $L_{\rm X,u} \approx 10^{29}-10^{30}$ erg s$^{-1}$. EU\,Cnc, the
CV in M\,67 \citep{vdbergea04}, and EF\,Eri \citep{schwea07} are two
such cases.  While polars make up a relatively small fraction of known
CVs (about 14\% in the catalog by \cite{rittkolb12}), their intrinsic
contribution to less-biased CV samples is not well known.  We estimate
that a small ($\sim$15\%) fraction of dwarf novae and an unknown
fraction of polars could have gone undetected by our {\em Chandra}
observation.  \nocite{heinea08}

The detection limit of the S03 catalog is $V\approx24$, which
corresponds to $M_V\approx10.5$ at the distance of NGC\,6791, taking
into account extinction. We compiled a sample of 27 CVs with parallax
distances from \cite{pattea08}, \cite{thor03}, \cite{harrea04}, and
\cite{thorea08}. Using the typical magnitude of each system (MAG1 in
\cite{rittkolb12}), we find that the corresponding $M_V$ distribution
extends down to $M_V\approx12.2$, with about 25\% of sources
(including the X-ray faint systems GW\,Lib and WZ\,Sge) below the S03
detection limit.

While the above arguments suggest that we should have detected most
CVs in NGC\,6791, \cite{gaenea09} found that the period distribution
of Sloan Digitized Sky Survey CVs has a spike around the period
minimum (80--86 min) corresponding to low-accretion rate systems. The
incompleteness of our observations to such faint systems, which have
an average magnitude of $M_g=11.6\pm0.7$, is significant.

\subsection{Active binaries in NGC\,6791} \label{disc_ab}

For ABs inside the half-mass radius of NGC\,6791, the {\em Chandra}
observation has a detection limit of about $(0.7-1) \times 10^{30}$
erg s$^{-1}$ (0.3--7 keV) if their coronal temperatures are between 1
and 2 keV. By comparison to the luminosity distribution of ABs in
M\,67, this implies that many more ABs can be present in NGC\,6791
than those listed in Table~\ref{tab_mem}.

\begin{figure}
\centerline{\includegraphics[width=8cm]{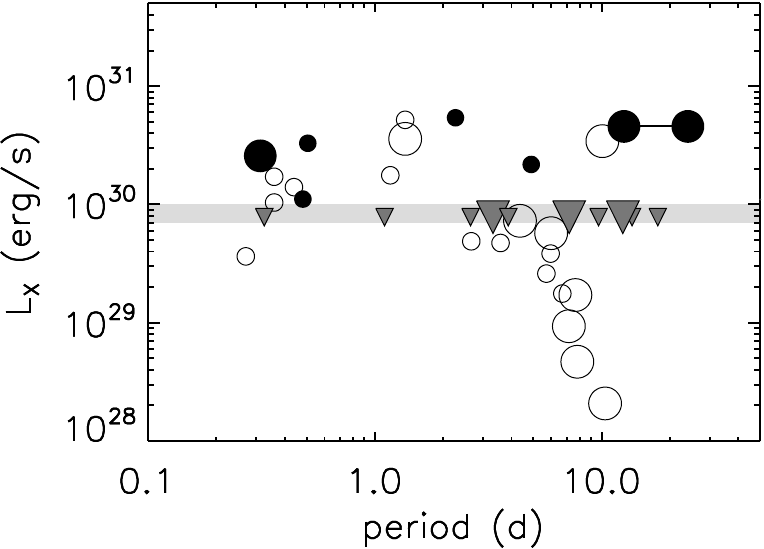}}
\caption{$L_{\rm X}$ versus orbital-period distribution of ABs inside
  the half-mass radii of M\,67 and NGC\,6791. Small symbols are
  systems on the main-sequence, larger symbols mark systems at the
  turnoff or on the sub-giant branch. The eccentric binaries S\,1242
  and S\,1272 in M\,67 are plotted at their pseudo-synchronous
  rotation periods. The two possible periods for CX\,23
  (Sect.~\ref{sec_ab}) are connected with a horizontal line. Our
  sample of ABs in NGC\,6791 (filled circles) is expected to be most
  incomplete for periods between a few days up to the circularization
  cutoff period, which agrees with the periods of the undetected
  periodic photometric variables in the cluster (filled
  triangles). The shaded area marks the upper limit for detection in
  NGC\,6791. We have used the recent proper-motion membership study
  from \cite{yadaea08} to re-assess membership for the M\,67 {\em
    Chandra} sources and thus classified three more ABs with known
  periods as likely cluster members; these are CX\,15, CX\,58 and
  CX\,61 (see Tables~2 and 3 in \cite{vdbergea04}). \label{fig_abs}}
\end{figure}

Fig.~\ref{fig_abs} shows the $L_{\rm X}$ versus orbital-period
distribution for ABs in M\,67 and NGC\,6791. The data for M\,67 (open
circles) were taken from \cite{vdbergea04}, but, for easier comparison
with the present analysis, the X-ray fluxes were recomputed such that
they correspond to emission from a 2-keV plasma. The ABs in M\,67 show
a dominant trend of decreasing $L_{\rm X}$ with increasing period for
orbital periods between about 4--5 and 10 days. This can be explained
by the activity-rotation relationship: stars in tidally-synchronized
binaries with longer orbital periods rotate slower, and therefore are
less active in X-rays, than stars in systems with shorter periods. For
the shortest-period ABs, $L_{\rm X}$ goes down again, which is a
result of saturated activity levels. The X-ray luminosity of
coronally-active stars does not exceed 0.001 times the bolometric
luminosity (e.g.~\cite{gued04}), which goes down for the
shortest-period binaries. For contact binaries, the maximum X-ray
luminosity is even lower, an effect known as super-saturation. In
Fig.~\ref{fig_abs}, stars that have evolved past the main-sequence
turnoff or that are at the turnoff are plotted with larger symbols. In
both clusters, these particular stars are only moderately evolved,
with estimated radii no larger than $\sim$3 $R_{\odot}$
\citep{flow96}, and all lie on or near the sub-giant branch. Except
for two cases, the sub-giant ABs do not show a different trend than
the main-sequence ABs (small symbols). The exceptions CX\,23 (V\,100)
in NGC\,6791 and S\,999 in M\,67 (upper right in Fig.~\ref{fig_abs})
lie above the overall rotation-activity trend, and are good targets
for follow-up study.

The ABs in M\,67 show a clear separation in orbital periods between
detections and non-detections (not shown in Fig.~\ref{fig_abs}, but
see \cite{vdbergea04}): practically all binaries with periods below
the tidal-circularization cutoff period (12 d in M\,67,
\citealt{lathmathea}) are detected and, except for eccentric systems
with pseudo-synchronous rotation periods shorter than 12 d, all normal
main-sequence and (sub)giant binaries with longer periods remain
undetected down to $L_{\rm X} \approx (2-6) \times 10^{28}$ erg
s$^{-1}$. This can be understood if the activity levels of the stars
in these wider systems are closer to those of single stars in the
cluster because their rotation is not (sufficiently) enhanced by tidal
coupling. If the ABs in NGC\,6791 from Table~\ref{tab_mem} show a
similar trend as those in M\,67, it is clear from Fig.~\ref{fig_abs}
that our {\em Chandra} observation of NGC\,6791 can only see the
brightest part of the AB population, and that our AB sample is most
incomplete for periods ranging from a few days up to the
circularization cutoff period in NGC\,6791, which is at least 15
d. This is consistent with the period distribution of the detected ABs
(filled circles) and non-detected periodic photometric variables with
membership propability $p_{\mu}>50$\% (filled triangles at the upper
limit of detection in Fig.~\ref{fig_abs}).  Variables for which the
period is ambiguous are excluded from the figure, and so is the
periodic variable V\,78 which is more likely to be a pulsator rather
than a binary \citep{kalu03}. Not all ABs are necessarily
photometrically variable all the time; a comprehensive radial-velocity
survey or deeper X-ray data would be useful to uncover more of them.

\subsection{On the origin of sub-subgiants} \label{disc_ssg}

Sub-subgiants in star clusters are typically detected in X-rays, but
the X-ray emission is explained by distinctly different types of
binary interaction, making it a rather inhomogenous group. Some
resemble ABs (see \cite{mathea03} for M\,67, and \cite{albrea01} and
\cite{heinea05b} for 47\,Tuc), others are CVs (AKO\,9 and PC1-V11 in
47\,Tuc, \citealt{albrea01}), and at least one but possibly two (in
NGC\,6397) are in binary MSPs where the X-rays originate in the shock
region between the mass outflow from the secondary and the energetic
pulsar wind \citep{bogdea10}. The disturbance of hydrostatic
equilibrium in the donor in response to mass transfer could explain
the CMD location for the CVs, but the sub-subgiants in M\,67 and
NGC\,6791 show no hint of ongoing mass transfer. Moreover, as has been
noted by \cite{mathea03}, the eccentric orbit of S\,1063 argues
against mass transfer in the past, as a (nearly) Roche-lobe filling
star would have quickly circularized the orbit on a time scale of
$\sim$10$^5$ yr. The fact that we see at least five sub-subgiants in
NGC\,6791 is an argument against invoking {\em any} short-lived
perturbed state (resulting from either regular binary evolution or
dynamical encounters with other cluster stars), and is more suggestive
of a hitherto overlooked binary evolutionary path. N-body simulations
of M\,67 by \cite{hurlea05} created one star below the subgiant branch
that offers a possible clue. This simulated star formed when two stars
in a binary merged after the donor ascended the sub-giant branch which
resulted in a common-envelope phase. The merger product is a single
giant with an undermassive core, which explains its location below the
normal cluster sub-giant branch. If this scenario applies to the
observed sub-subgiants, which all show signs of binarity, the systems
should originally be in triples. No primordial triples were included
in the models by \cite{hurlea05}, but understanding the origin of
sub-subgiants would be a strong motivation for taking them into
consideration in future cluster simulations. As has been discussed in
the context of blue stragglers by \cite{perefabr09}, triples can
indeed facilitate the formation of very close inner binaries that
eventually merge or enter a phase of mass transfer.

\subsection{X-ray source populations of old open clusters} \label{disc_oc}

We compare the close-binary populations in three old open
clusters---NGC\,6791, M\,67, NGC\,6819--- as probed by X-ray
observations to look for trends with cluster properties, specifically
mass or age. NGC\,6819 was recently studied in X-rays by
\cite{gosnea12} using {\em XMM-Newton} data. As we discuss in more
detail below, compared to M\,67 and NGC\,6791, classification of the
X-ray sources in the field of NGC\,6819 is far less complete at this
point, but we include the cluster as it extends the age range of the
sample down to 2.4 Gyr. Previous studies have shown that all three
clusters have a high fraction of binaries: almost 40\% in NGC\,6819
(spectroscopic solar-type binaries; \citealt{math08}), $\sim$50\% in
M\,67, and $\gtrsim$25-30\% in NGC\,6791 (photometric estimates from
the width of the main sequence; \citealt{fanea} and
\citealt{bediea08}). In order to set a uniform sensitivity, we only
consider X-ray sources brighter than $L_{\rm X} = 1 \times 10^{30}$
erg s$^{-1}$. We do not include clusters that have been observed with
{\em ROSAT} only (see \cite{bell97} for an overview) as these
observations are generally shallower, and the larger positional
uncertainties result in less secure source identifications.

Table~\ref{tab_oc} gives an overview of the cluster properties and the
number of X-ray sources $N_X$, in total and separated by class,
detected within the approximate half-mass radii. For NGC\,6791 we
refer to the list of sources (members and possible members) in
Table~\ref{tab_mem}. \cite{platea11} estimate a lower limit to the
cluster mass of $\sim$5000 $M_{\odot}$; taking into account detection
incompleteness, stellar remnants, and binarity, the total mass is
likely not higher than $\sim$7000 $M_{\odot}$. For M\,67 we adopt a
total mass of $\sim$1100 $M_{\odot}$ \citep{richea}. \cite{fanea}
measured the half-mass radii for several types of stars (giants,
single main-sequence stars, and main-sequence binaries); as a result
of the effects of mass segregation, they find values ranging from
$\sim$7\farcm2 to 12\arcmin, depending on the typical mass of the
objects.  For simplicity, we count the number of {\em Chandra} sources
inside the ACIS-I field (16\farcm8 $\times$ 16\farcm8), which is close
to the area inside any of the $r_h$ estimates by \cite{fanea}. For
both M\,67 and NGC\,6791 we do not count unidentified sources, or
unclassified sources with candidate counterparts that lie away from
the main sequence, as possible members because the extra-galactic
$\log N - \log S$ curves suggest they are most likely background AGN.

\cite{gosnea12} detect twelve sources inside $r_h=3$\farcm3 down to
$L_{\rm X} = 1\times10^{30}$ erg s$^{-1}$ (0.2--10 keV) in a {\em
  XMM-Newton} pointing of NGC\,6819. The estimated number of
background sources suggests at most six to seven of these are cluster
members. The five sources with candidate optical/UV counterparts
include potentially interesting systems like a candidate CV, a
candidate qLMXB (the nature of both need to be confirmed
spectroscopically), a possible sub-subgiant (X\,9), and an RS\,CVn
binary that may have been formed in a dynamical encounter (X\,6).
Given the depth of the optical catalog ($V\approx 24$) used to look
for matches, it is reasonable to assume that an upper limit to the
number of cluster ABs is the number of sources with a candidate
counterpart on the (binary) main sequence or (sub-)giant branch that
cannot be excluded as a non-member. There is at most one such system
(X\,6). Due to the incomplete source classifications, the total
cluster X-ray luminosity is highly uncertain, and we do not list it in
Table~\ref{tab_oc}. If we simply take the sum of the luminosities of
the twelve sources and divide it by two, we find $\log (2 L_{30}/M)
\approx29.1$ (0.2--10 keV); as this estimate is already very
approximate, we do not attempt to convert this value to the 0.3--7 keV
band. \cite{kaliea01} estimate the mass of NGC\,6819 to be
$\sim$2600$M_{\odot}$.

The numbers of CVs ($N_{X,CV}$) and sub-subgiants ($N_{X,S}$) are
small\footnote{The CV EU\,Cnc in M\,67 is fainter in X-rays, and
  strictly speaking it has not been demonstrated that it is a cluster
  member. The well-studied sub-subgiant S\,1113 in M\,67 is not
  counted in Table~\ref{tab_oc} as it lies 13\farcm7 away from the
  cluster center. Likewise, the sub-subgiants CX\,68 and CX\,77 in
  NGC\,6791 lie at $r>r_h$.}; only in NGC\,6791 are there any
spectroscopically confirmed CVs above our luminosity cutoff. A
positive correlation of $N_{X,CV}$ and $N_{X,S}$ with cluster mass is
allowed. Based on a comparison with the field CV space density, we
already argued in Sect.~\ref{disc_cv} that the CVs in NGC\,6791 could
be primordial, and the same is true for the (candidate) CVs in M\,67
and NGC\,6819. This indeed implies that $N_{X,CV}$ is proportional
to cluster mass.

The value of $N_{X,AB}$ in Table~\ref{tab_oc} gives the number of
normal ABs, i.e.~it does not include blue-straggler systems, which may
have a non-standard evolutionary history, or the long-period AB-like
sources (such as the yellow stragglers S\,1040 and S\,1072 in M\,67;
\citealt{vdbergea04}). Comparison of $N_{X,AB}$ for the three clusters
shows that a simple scaling with the present-day cluster mass is
excluded. This is true for the ABs with evolved components
($N_{X,AB-G}$), which have wider orbits and may thus be easier
involved in dynamical interactions, but also for ABs on the main
sequence ($N_{X,AB-MS}$). Although NGC\,6791 is 4.5--6.4 times more
massive than M\,67, it has 0.9--1.6 times the number of ABs. NGC\,6819
has 2.4 times the mass of M\,67, but only has a fraction ($\sim$0.13)
of the number of ABs. These ratios show that $N_{X,AB}$ normalized by
cluster mass does not show a trend with age or mass either.  M\,67 has
the highest specific AB abundance ($\sim$0.014 per half the cluster
mass), while this number is at least $\sim$3 times lower in NGC\,6819
and NGC\,6791, which are both more massive than M\,67, but younger and
older, respectively

It is unclear if the relative dearth of ABs in NGC\,6819 and NGC\,6791
with respect to M\,67 can be explained by stellar or binary-evolution
arguments, or if cluster dynamics plays a role. Possibly, M\,67 is an
outlier. Several studies have found that this cluster is dynamically
highly evolved \cite[e.g.][]{davesand10}. If, as a result of mass
segregation, it lost a significant fraction of its mass due to the
evaporation of preferentially low-mass stars, the current binary
population would appear as representive of a much more massive
cluster. To explain the discrepancy in the number of ABs, this effect
should be more significant for M\,67 than for NGC\,6819 and
NGC\,6791. Another difference is that the X-ray source population of
M\,67 contains binaries (blue and yellow stragglers) with periods in
excess of hundreds of days that look like coronal emitters (but note
that these are not included in $N_{X,AB}$ in Table~\ref{tab_oc}); such
puzzling systems are not seen in the other two clusters. Clearly, a
larger cluster sample spanning wider ranges in age and mass needs to
be studied to make sense of what is the norm and what the exception,
and understand the X-ray emission of old open clusters (to which ABs
contribute an appreciable fraction\footnote{$\sim$15\% in NGC\,6791,
  $\sim$45\% in M\,67}) in general. As NGC\,6791 is one of the oldest
open clusters known, we expand our comparison to globular clusters in
the next section.

There are two more old ($\gtrsim 2$ Gyr) open clusters that have been
observed in X-rays, but that we have not considered here. The young
(1.7--2 Gyr) and nearby cluster NGC\,752 is spatially extended with
members spread over a much larger region ($\sim$90\arcmin~in diameter)
than covered by the {\em Chandra} and {\em XMM-Newton} pointings. The
census of X-ray sources could thus be very incomplete. We note that
the list of detected members from \cite{giarea08} includes just one
source with $L_{\rm X} \gtrsim 1\times10^{30}$ erg s$^{-1}$, which is
matched to a likely main-sequence AB. \cite{gond05} analyzed {\em
  XMM-Newton} data of NGC\,188 with a sensitivity of $L_{\rm X}=9
\times 10^{29}$ erg s$^{-1}$ (0.5--2 keV), about six times deeper than
the ROSAT study by \cite{bellea}. However, the chosen aimpoint is
offset from the cluster center by almost 6\arcmin. Only new
observations can tell whether this explains why no sources are
detected in 1/4--1/3 of the contiguous area inside the half-mass
radius that is farthest from the observation aimpoint. In any case,
this unfortunate circumstance makes NGC\,188 less suitable to be
included in our comparison.\footnote{Moreover, some sources seem to be
  missing from the list of six detected cluster members in
  \cite{gond05}. ROSAT source X\,29 in \cite{bellea}, which is the
  brightest cluster X-ray source, is not included in the source list
  despite being clearly visible in the EPIC-MOS images; in the {\em
    XMM-Newton} Serendipitous Source Catalog \citep{watsea09} it has a
  maximum source detection likelihood of 279. ROSAT source X\,21 is
  identified by \cite{bellea} with a likely short-period AB and
  cluster member. The offset between this optical source and the
  corresponding {\em XMM-Newton} source S\,17 is 3.2\arcsec, and it is
  unclear why it was not identified as a potential counterpart given
  the match radius of 8\arcsec~that is adopted by \cite{gond05}.}

\begin{table*}
\begin{center}
\caption{X-ray sources in globular and old open clusters ($r \lesssim r_h$,
  $L_{\rm X}\gtrsim1\times10^{30}$ erg s$^{-1}$) \label{tab_oc}}

\begin{tabular}{lccccccccc}
\hline
\hline
cluster     & age        & $M$              & $N_X$      & $N_{X, CV}$  & $N_{X, S}$ & $N_{X, AB}$ & $N_{X,AB-MS}$ & $N_{X,AB-G}$  & $\log (2~L_{30}$/$M$) \\
            & (Gyr)      & ($M_{\odot}$)     &            &             &          &       &     &        &         \\
\hline
NGC\,6819   &  2--2.4    & 2600             & {\em 6--7} &  {\em 1?}  &  {\em 1?} &   {\em 1?} & {\em 0} & {\em 1?} &   \ldots            \\ 
M\,67       &  4         & 1100             & 12         &    0       & 1         &  7--8      & 5--6 & 2 &  28.9        \\ 
NGC\,6791   &  8         & 5000--7000       & 15--19     &   3--4     & 3         & 7--11      & 5--9 & 2  &  28.6--28.8           \\  
\hline
47\,Tuc     & 11.2       & $1.3\times10^{6}$ & $\sim$200 & 30[--119]    & 10   & 42[--131]   & 34 & 8 & 28.0 \\
NGC\,6397   & 13.9       & $2.5\times10^{5}$ & 15--18    &  11        & 2          & 0--2      & 0--2 & 0 & 27.7\\

\hline
\end{tabular}
\tablecomments{We list the number of sources identified with cluster
  members ($N_X$), CVs ($N_{X,CV}$), sub-subgiants ($N_{X,S}$), and
  ABs ($N_{X,AB}$). We sub-divide the ABs into those on the main
  sequence ($N_{X,AB-MS}$) and evolved systems ($N_{X,AB-G}$). The
  numbers for NGC\,6819, printed in {\em italics}, are especially
  uncertain due to the limited optical follow up. Log (2 $L_{30}$/$M$)
  is, in log units, the ratio of the total X-ray luminosity of the
  $N_X$ sources with $L_{\rm X}\gtrsim 1\times 10^{30}$ erg s$^{-1}$,
  to the cluster mass ($M$) divided by two to account for the
  selection of sources inside the half-mass radius. The X-ray
  luminosity refers to the 0.3--7 keV band and corresponds to emission
  from a 2 keV plasma. The horizontal line separates the open clusters
  from the globular clusters. References: \cite{gosnea12} for
  NGC\,6819, \cite{vdbergea04} for M\,67, \cite{bogdea10} and
  \cite{cohnea10} for NGC\,6397, \cite{heinea05b} for 47\,Tuc. }
\end{center}
\end{table*}

\subsection{Comparison with globular clusters}

While almost 80 globular clusters have been studied with {\em
  Chandra}, only a handful have been observed deep enough to uncover a
significant fraction of the ABs. Here we focus on two globular
clusters for which the {\em Chandra} sensitivity reaches well below
the luminosity cutoff adopted in the previous section. The nearby
core-collapsed cluster NGC\,6397 was observed down to a limiting
luminosity of $\sim$$9\times10^{28}$ erg s$^{-1}$ \citep{bogdea10}. We
adopt a cluster mass of $2.5\times10^{5}$ $M_{\odot}$
\citep{pryomeyl93}. \cite{cohnea10} classified most of the 79 {\em
  Chandra} sources inside $r_h=$ 2\farcm33 from \cite{bogdea10} using
      {\em HST} data, and found numerous CVs and ABs, and two
      sub-subgiants (at least one is in a binary with a MSP). About
      20\% of the sources are brighter than our cutoff value; these
      are mostly CVs or neutron-star binaries. While there are plenty
      of X-ray faint ABs, there are just a few at most above
      $1\times10^{30}$ erg s$^{-1}$; the range listed in
      Table~\ref{tab_oc} is zero to two, but those two systems have
      very red colors and are unlikely cluster members. 47\,Tuc is a
      massive cluster ($1.3\times10^{6}$ $M_{\odot}$;
      \citealt{pryomeyl93}).  We estimate that within $r_h=$
      2\farcm79, about 28 of the 232 sources bright enough to be
      included in our selection are background AGN, leaving $\sim$200
      sources associated with the cluster. As of yet, only part of
      these have been classified
      \citep{edmogillea03a,edmogillea03b}. We make a very conservative
      estimate of the number of CVs and ABs. A lower limit is derived
      by counting the number of (candidate) CVs and ABs in Table 2 of
      \cite{heinea05b}, while an upper limit (given in square brackets
      in Table~\ref{tab_oc}) is estimated by adding all unidentified
      and unclassified sources to either the CVs or ABs and
      subtracting the estimated contribution from AGN. Sub-subgiants
      are more conspicious, so their number is less uncertain. To
      compute the cluster X-ray luminosity, we sum the luminosities
      for the $N_X$ individual sources, and subtract an approximate
      contribution for the background AGN based on the mean luminosity
      of the unidentified sources. For both clusters we converted the
      fluxes listed in \cite{bogdea10} and \cite{heinea05b} to the
      0.3--7 keV band using PIMMS, adopting a 2-keV plasma model.

Table~\ref{tab_oc} readily shows that the total number of ABs above
$L_{\rm X}=1 \times 10^{30}$ erg s$^{-1}$, normalized by cluster mass,
is much lower in the two globular clusters than in M\,67 and
NGC\,6791. For 47\,Tuc the difference is a factor of 65--230 compared
to M\,67, whereas NGC\,6397 likely does not have any ABs at all above
this luminosity. Lowering the X-ray luminosity cutoff value would
increase the numbers of ABs especially for NGC\,6397, but this is not
enough to make up for the lower specific X-ray luminosity. The poor
statistics in the open clusters left aside, CVs appear
underrepresented in globular clusters as well, by about an order of
magnitude. A likely explanation is that dynamical encounters have a
net effect of destroying ABs---{\em both} those with evolved
components and wider orbits {\em and} main-sequence systems---and
CVs. At the same time, the correlation with encounter rate found by
\cite{poolhut06} implies that at least CVs are formed as well; perhaps
this is why the specific frequencies in open and globular clusters are
less disparate for CVs than for ABs. AB-like sub-subgiants are
underabundant in globular clusters as well, but on the other hand
sub-subgiants in (candidate) binary MSPs are not found in open
clusters (note that in Table~\ref{tab_oc} the different types of
sub-subgiants are grouped together because some have not been studied
well enough to make the distinction).

The relative lack of ABs in NGC\,6397 with respect to other globular
clusters was already noted by \cite{bassea04} and \cite{grin06}, who
ascribed it to its post-collapsed state. We now show that 47\,Tuc is
deficient in ABs, too. M\,4, studied by \cite{bassea04} and not
included in Table~\ref{tab_oc}, shows the same effect. Even if all
fifteen sources or so that are brighter than $1\times10^{30}$ erg
s$^{-1}$ were ABs (which is not the case as there are also CVs,
background and foreground sources, and an MSP among them), it is not
enough to reach the number of ABs expected based on scaling by mass
from open clusters. \cite{miloea12} find a lower overall binary
fraction in globulars compared to open clusters \citep{sollea10}. If
this is the result of binary destruction, the X-ray emitting, and thus
hardest, binaries are affected as well. We cannot exclude that other
factors are important. There is a trend of decreasing AB frequency
with age for all clusters in Table~\ref{tab_oc} except NGC\,6819.

In \cite{verb00} it was first pointed out that the total {\em ROSAT}
luminosity per unit mass of most globular clusters is lower than that
of M\,67.  While M\,67 may be exceptional in some ways (see
Sect.~\ref{disc_oc}), the last column of Table~\ref{tab_oc} shows that
the specific X-ray luminosity of globular clusters may be lower than
for open clusters in general. CVs, ABs, and sub-subgiants are all
responsible for this. The X-ray emission from qLMXBs and MSPs in
globular clusters, of which there are no confirmed cases in old open
clusters, cannot make up for the lack of other types of faint
sources. A more extensive investigation, based on a larger sample of
globular clusters and old populations, is done by Heinke et al.~(in
preparation) to investigate the effects of age, central density, and
metallicity on differences in the X-ray emissivity.

\section{Conclusions} \label{sec_conc}

We have studied NGC\,6791, which is one of the oldest open clusters
known, in X-rays for the first time. Of the twenty {\em Chandra}
sources that are identified with proper-motion cluster members,
nineteen are binaries. We find a mix of CVs, ABs, and sub-subgiant
binaries. With optical follow-up spectroscopy we confirm the
classification of one CV in the cluster, and identify one new
candidate CV. The origin of these and the other two cluster CVs is
likely primordial. A comparison with the X-ray sources in the younger
cluster NGC\,6819 suggests that in order to find numerous
(main-sequence) ABs brighter than $L_{\rm X} = 1\times10^{30}$ erg
s$^{-1}$ in open clusters, one must look at those older than $\sim$2.5
Gyr.  The mass-specific frequency of main-sequence and sub-giant ABs
in NGC\,6791 (8 Gyr) is 3--7 times lower than in M\,67 (4 Gyr); it is
not clear why this is the case. Comparison with a small number of
globular clusters shows that all three source classes that are mainly
responsible for the X-ray emission from old open clusters are
under-represented in the globulars. This accounts for the lower total
X-ray luminosity per unit mass of globular clusters, and indicates
that the net effect of dynamical encounters may be destruction
of--even the hardest--binaries. The cutoff luminosity for comparing
source populations among clusters was set at $L_{\rm X} =
1\times10^{30}$ erg s$^{-1}$ (0.3--7 keV). A lower cutoff luminosity
might be needed to account for effects of variations of the AB X-ray
luminosity function with age or metallicity. While \cite{verb00} note
that population-II ABs may have lower X-ray luminosities, it remains
unclear in this context why NGC\,6791, which has a super-solar
metallicity, should also have relatively fewer ABs above our limiting
$L_{\rm X}$. We defer a more thorough investigation to a future
study. Deeper data are needed for NGC\,6791 to match the already deep
existing data for M\,67, 47\,Tuc, and NGC\,6397. The X-ray derived AB
populations may then be compared to those derived from other
wavelengths, e.g.~optical variability studies. Also in this respect,
studies of NGC\,6791 may benefit from being in the field of view of
the {\em Kepler} satellite.

\begin{acknowledgments}
The authors would like to thank J.~Hong for providing the routines to
compute energy quantiles, C.~Heinke for comments on the manuscript,
and K.~Cudworth for providing a catalog of proper motions. This
research was supported by {\em Chandra} grant GO0-11110X. Some of the
observations reported here were obtained at the MMT Observatory, a
joint facility of the Smithsonian Institution and the University of
Arizona.
\end{acknowledgments}

{\it Facilities:} \facility{CXO}, \facility{FLWO:1.5m (FAST)}, \facility{MMT (Hectospec)}

\end{document}